\documentstyle[12pt,graphics]{article}
\newcommand{\A}{\alpha}
\newcommand{\B}{\beta}
\newcommand{\D}{\delta}
\newcommand{\DE}{\Delta}
\newcommand{\G}{\gamma}

\newcommand{\eps}{\epsilon}

\newcommand{\bea}{\begin{eqnarray}}
\newcommand{\be}{\begin{equation}}
\newcommand{\ee}{\end{equation}}
\newcommand{\ena}{\end{eqnarray}}
\newcommand{\beano}{\begin{eqnarray*}}
\newcommand{\enano}{\end{eqnarray*}}
\newcommand{\hf}{\frac{1}{2}}

\newcommand{\sthalf}{{\textstyle {3\over2}}}
\newcommand{\ad}{\mbox{ad}}

\newcommand{\tr}{\mbox{\hs{2}tr\hs{2}}}

\newcommand{\nn}{\nonumber \\ }
\newcommand{\pa}{\partial}

\newcommand{\vs}[1]{\rule[- #1 mm]{0mm}{#1 mm}}
\newcommand{\hs}[1]{\hspace{#1 mm}}

\newcommand{\mb}[1]{\hs{5}\mbox{#1}\hs{5}}

\newcommand{\WZW}{{\rm WZW}}
\newcommand{\del}{\partial}
\newcommand{\cg}{{\mbox{$\cal{G}$}}}
\newcommand{\ch}{{\mbox{$\cal{H}$}}}
\newcommand{\ck}{{\mbox{$\cal{K}$}}}

\newcommand{\cs}{\mbox{$\cal{S}$}}

\newcommand{\cw}{\mbox{$\cal{W}$}}
\newcommand{\cz}{\mbox{$\cal{Z}$}}

\newcommand{\wgs}{{\cal W}^{\cal G}_{\cal S}}

\font\fld=msbm10 at 12 pt

\newcommand{\fl}[1]{\mbox{\fld #1}}     
\newcommand{\NPB}[1]{{\it Nucl.~Phys.}~{\bf B#1}}
\newcommand{\PLB}[1]{{\it Phys.~Lett.}~{\bf B#1}}

\newcommand{\CMP}[1]{{\it Comm.~Math.~Phys.}~{\bf #1}}

\newcommand{\PL}[1]{Phys.\ Lett.\ {\bf #1}}

\newcommand{\PR}[1]{Phys.\ Rev.\ {\bf #1}}

\newtheorem{defi}{Definition}
\newtheorem{theo}{Theorem}

\newtheorem{cor}{Corollary}
\newcommand{\li}{
\begin{picture}(38,12)
\linethickness{1.4pt}
\put(2,6){\line(1,0){38}}
\end{picture}}
\newcommand{\dli}{
\begin{picture}(38,12)
\linethickness{1.4pt}
\put(2,4){\line(1,0){38}}
\put(2,8){\line(1,0){38}}
\end{picture}}
\newcommand{\tli}{
\begin{picture}(38,12)
\linethickness{1pt}
\put(2,3){\line(1,0){38}}
\put(2,6){\line(1,0){38}}
\put(2,9){\line(1,0){38}}
\end{picture}}
\newcommand{\ci}{
\begin{picture}(8,12)
\put(0,0){\Huge $\circ$}
\end{picture}}
\newcommand{\bu}{
\begin{picture}(8,12)
\put(0,0){\Huge $\bullet$}
\end{picture}}
\newcommand{\cir}[1]{
\begin{picture}(8,26)
\put(2,14){#1}
\put(0,0){\Huge $\circ$}
\end{picture}}
\newcommand{\bul}[1]{
\begin{picture}(8,26)
\put(2,14){#1}
\put(0,0){\Huge $\bullet$}
\end{picture}}
\newcommand{\up}[2]{
\begin{picture}(8,50)
\put(9,14){#1}
\put(0,0){\Huge $\circ$}
\linethickness{1.4pt}
\put(6,10){\line(0,1){29}}
\put(13,41){#2}
\put(0,37){\Huge $\circ$}
\end{picture}}
\newcommand{\hli}{
\begin{picture}(27,12)
\linethickness{1.4pt}
\put(2,6){\line(1,0){10}}
\put(15,6){\line(1,0){4}}
\put(22,6){\line(1,0){2}}
\end{picture}}
\newcommand{\lih}{
\begin{picture}(25,12)
\linethickness{1.4pt}
\put(5,6){\line(1,0){2}}
\put(10,6){\line(1,0){4}}
\put(17,6){\line(1,0){10}}
\end{picture}}
\newcommand{\hlih}{\hli\lih}
\newcommand{\Aone}{\begin{picture}(20,20)\put(0,1){\ci}\end{picture}}
\newcommand{\Aonex}{\begin{picture}(20,20)\put(0,1){\bu}\end{picture}}
\newcommand{\twoAone}{\begin{picture}(50,20)
\put(0,1){\ci}\put(30,1){\ci}\end{picture}}
\newcommand{\twoAonex}{\begin{picture}(50,20)
\put(0,1){\ci}\put(30,1){\bu}\end{picture}}
\newcommand{\Atwo}
{\begin{picture}(70,20)
\put(0,1){\ci}
\put(8,1){\li}
\put(46,1){\ci}\end{picture}}
\newcommand{\Ctwo}{\begin{picture}(70,20)
\put(0,1){\ci}
\put(8,1){\dli}
\put(46,1){\bu}
\end{picture}}
\newcommand{\Gtwo}{\begin{picture}(70,20)
\put(0,1){\ci}
\put(8,1){\tli}
\put(46,1){\bu}
\end{picture}}
\newcommand{\Atwoe}{\begin{picture}(80,35)
\put(0,1){\ci}
\put(8,1){\li}
\put(46,1){\ci}
\put(14,10){\line(1,1){17}}
\put(25,22){\ci}
\put(39,27){\line(1,-1){16}}
\end{picture}}
\newcommand{\Ctwoe}{\begin{picture}(130,20)
\put(0,1){\ci}
\put(8,1){\dli}
\put(46,1){\bu}
\put(54,1){\dli}
\put(92,1){\ci}
\end{picture}}
\newcommand{\Gtwoe}{\begin{picture}(130,20)
\put(0,1){\ci}
\put(8,1){\li}
\put(46,1){\ci}
\put(54,1){\tli}
\put(92,1){\bu}
\end{picture}}
\begin{document}
\renewcommand{\thefootnote}{\fnsymbol{footnote}}
\newpage
\pagestyle{empty}
\setcounter{page}{0}

\vs{10}
\begin{center}

{\Large {\sc \protect{On the Classification of Real Forms} \\
\protect{of Non-Abelian Toda Theories} \\[7pt]
\protect{and \cw-algebras}}}\\[1cm]

\vs{2}

{\large \protect{J.M. Evans\footnote{e-mail:
J.M.Evans@damtp.cam.ac.uk}}}, {\em DAMTP, University of
Cambridge\footnote{Silver Street, Cambridge CB3 9EW, UK.}} \\[2mm]
{\large and} \\[2mm] {\large \protect{J.O. Madsen\footnote{e-mail:
madsen@fpaxp1.usc.es}}}, \protect{ {\em Universidade de Santiago de
Compostela}\footnote{Dpto de F{\'\i}sica de Part{\'\i}culas, 
E-15706 Santiago de Compostela, Spain}}

\end{center}
\vfill

\centerline{ {\bf Abstract}}

\noindent
We consider conformal non-Abelian Toda theories obtained
by hamiltonian reduction from Wess-Zumino-Witten models based on 
general real Lie groups. 
We study in detail the possible choices of reality conditions 
which can be imposed on the WZW or Toda fields and prove
correspondences 
with $sl(2,\fl{R})$ embeddings into real Lie algebras and with  
the possible real forms of the associated $\cw$-algebras.
We devise a a method for finding all real embeddings which can be obtained 
from a given embedding of $sl(2,\fl{C})$ into a complex Lie algebra.
We then apply this to give a complete classification of real
embeddings which are principal in some simple 
regular subalgebra of a classical Lie algebra.

\vfill
\rightline{hep-th/9802201}
\rightline{US-FT/2-98}
\rightline{DAMTP/97-148}
\rightline{December 1997}

\newpage
\pagestyle{plain}
\setcounter{footnote}{0}
\renewcommand{\thefootnote}{\arabic{footnote}}
\section{Introduction} 
\label{sec-0}

Toda field theories provide a rich set of examples of integrable models 
in two dimensions which can be of either conformal or massive type.
In each case integrability rests on some underlying Lie algebra (or
superalgebra) which renders the theory tractable despite highly
non-trivial self-interactions. 
In this paper we shall consider conformally-invariant Toda theories \cite{LS}
which can be obtained by hamiltonian reduction from a Wess-Zumino-Witten 
(WZW) model \cite{ORetc},\cite{BTD},\cite{PhysRep}. 
More precisely, one can construct a Toda theory by taking
the lagrangian for a WZW model based on some real, non-compact
Lie group $G$ and gauging a suitable nilpotent subalgebra of its
Kac-Moody symmetry. 
This produces a new field theory in which a 
WZW lagrangian for some subgroup $G_0 \subset
G$ is obtained, though modified by the addition of a potential term of
a special kind. 
The resulting Toda theory is usually referred to as abelian or
non-abelian depending on the nature of
$G_0$ (the original group $G$ is always non-abelian).

Perhaps the most important aspect of conformally-invariant Toda models
is that they provide natural realizations of a certain class of 
infinite-dimensional chiral symmetry algebras which contain a 
Virasoro subalgebra. These are referred to generically as 
extended conformal algebras, or \cw-algebras.
The simplest examples are the  
superconformal algebras and the Kac-Moody/Virasoro semi-direct
products, which
have generators with spins at most two that close on themselves
in the conventional sense of Lie algebras.
In general, however, \cw-algebras may have generators of spin higher
than two and they are usually non-linear
in the sense that classical Poisson brackets or quantum commutators 
close only onto expressions that are polynomial in the generators.
The great interest in \cw-algebras stems from the fact 
that they greatly extend the
domain of {\em rational\/} models, in which 
the Hilbert space of the theory splits into finitely many 
irreducible representations of the chiral symmetry algebra.
For a comprehensive review see \cite{BS}.

In the hamiltonian reduction approach mentioned above, the gauging of a
nilpotent subgroup of the original WZW model 
corresponds to imposing a certain set of first-class
constraints on the Kac-Moody currents. 
It is precisely the gauge-invariant polynomials in these currents
which become generators of the \cw-symmetry in the Toda theory.
This allows one to understand rather easily the spin content of the 
\cw-algebra in terms of some simple group theory.
Moreover, the Toda point-of-view has proved crucial to the study 
of \cw-algebras, not just because it has provided Lagrangian realizations of 
known \cw-algebras, but also because it has lead to the
construction of many new examples which are more difficult to obtain
by other means.
It is a 
plausible conjecture that hamiltonian reduction can provide the basis
for a complete classification of \cw-algebras. 

Our aim in this paper is to carry out a detailed study of
hamiltonian reduction and the Toda theories and 
\cw-algebras which emerge when $G$ is a
general real Lie group.
Previously in the literature, attention has been confined 
largely to examples in which $G$ is maximally non-compact 
(i.e.~the {\em split\/} real form) and, 
although the necessary general apparatus has been in existence for some time,
no systematic analysis of other possibilities seems to have been
undertaken. As indicated in our earlier preliminary investigation
\cite{EM}, 
the consideration of Toda models based on general real forms turns out
to be interesting for a number of inter-related reasons.
\begin{itemize}

\item[-] Different choices of real form for $G$ yield different real
forms for the \cw-algebra which appears in the resulting Toda theory. 
This suggests a systematic way of generating new real forms of \cw-algebras
and of classifying them. 

\item[-] The existence of inequivalent real forms is clearly crucial 
as regards the representation theory of \cw-algebras.
We can compare with the case of ordinary Lie algebras, where 
the representation theory of compact and non-compact real forms 
is very different. 
Recently, this question has been raised for the simpler cases of {\em finite\/}
\cw-algebras \cite{BoHaTj} 
(finite \cw-algebras can be constructed from finite dimensional Lie
algebras in the same way the ``affine'' \cw-algebras are constructed
from affine Lie algebras). 

\item[-] The preceding remarks take on special significance 
when one bears in mind that the reduced theory often has a residual 
Kac-Moody symmetry corresponding to a subgroup $K \subset G$. 
Different real forms for $G$ will lead to 
inequivalent real forms of $K$. 
For many applications of Kac-Moody algebras 
we are most interested in the compact real form; we may therefore
expect that when we consider \cw-algebras it will be of great interest
to determine whether we are able to choose a real form of 
$G$ so as to make $K$ compact. 
\end{itemize}

The conformal reduction of a WZW model based on a
group $G$ can be phrased in terms of choosing an
$sl(2,\fl{R})$ subalgebra of the Lie algebra \cg .
Our approach can therefore be interpreted as a
systematic search for $sl(2,\fl{R})$ embeddings into real Lie 
algebras. We shall present a method for 
determining all possible real forms 
of a given Lie algebra compatible with the existence of 
a given $sl(2,\fl{R})$ subalgebra $\cs \subset \cg$, and we shall use
this method to classify a natural subset 
of the embeddings allowed when $\cg$ is classical. 
Note that the 
{\em compact} real form of $G$ is {\em never} compatible with hamiltonian
reduction in this sense, since there are no embeddings of 
$sl(2,\fl{R})$ into a compact real Lie algebra. 

It is also important to draw a clear distinction between what we are trying
to achieve here and the theory and classification of embeddings 
of $sl(2,\fl{C})$ into {\em complex} 
Lie algebras, as studied by Dynkin about 40 years ago \cite{Dy}.
His classic results can 
be transferred immediately to deal with 
embeddings of $sl(2,\fl{R})$ in {\em maximally 
non-compact} or {\em split} real Lie algebras but, 
as far as we know, very little has been done regarding embeddings of 
$sl(2,\fl{R})$ into general real Lie algebras, and no form of 
classification seems to have been carried out until now. 
This actually goes some way towards explaining why 
so little has been done in the area of hamiltonian reduction based on general 
real Lie groups, which we hope to rectify here. 

In a recent paper \cite{BiGu}, Bina and G\"unaydin have carried out the 
classification of real forms of simple non-linear superconformal
algebras, as well as others known as 
`quasi-superconformal' and `super-quasi-superconformal' or
$\fl{Z}_2\times\fl{Z}_2$ superconformal algebras \cite{FRS}, 
using an approach based on 
quaternionic and superquaternionic symmetric spaces of simple Lie groups and 
supergroups. Although we consider exclusively bosonic algebras in this
paper, we have already indicated in \cite{EM} how our methods
can be extended to the hamiltonian reduction 
of superalgebras, and it 
would be interesting to apply them to the special cases of superconformal and 
$\fl{Z}_2 \times \fl{Z}_2$ algebras 
discussed in \cite{BiGu}.
The quasi-superconformal algebras (such as the Bershadsky algebra 
$\cw_3^2$) involve hamiltonian reductions with half-integer gradings, 
which again falls outside our present aims.
Nevertheless our methods are in principle applicable to these cases too.

The paper is organized in the following way: In section \ref{sec-1} we start 
by giving a brief account of hamiltonian reduction as applied to
a WZW model based on a general real Lie group. In section \ref{sec-2} 
we recall some facts about the real forms of a complex Lie algebra 
before stating and proving some general results. 
These establish precise correspondences between real forms
of a Lie algebra consistent with an $sl(2, \fl{R})$ embedding, real forms 
of non-abelian Toda theories, and real forms 
of extended conformal algebras. 
In section \ref{sec-rank2} we illustrate our 
general results by discussing in some detail 
the reductions and real forms for integral embeddings 
in algebras of rank two.
In the following section \ref{sec-3} we further develop our 
these techniques to cope with algebras of arbitrary rank.
We apply them to classify all integral embeddings which are 
principal in some regular subalgebra of
one of the classical algebras of type 
$A_n,\,B_n,\,C_n,$ or $D_n$. 
In section \ref{sec-4} we give a summary of the 
classification, followed by a discussion of our results
and some suggestions for future work.
Two appendices contains some selected 
calculations. 

\section{Non-Abelian Toda Theories and Conformal 
\\ Hamiltonian Reduction.}
\label{sec-1}

In this section we give a brief review of the 
conformal 
hamiltonian reduction of a WZW model based on a general real Lie group
$G$ and how this gives rise to a Toda theory (non-abelian in general)
realizing a certain $\cw$-algebra. 
This is in principle a special case of the general reduction
schemes described in sections 2 and 3 of \cite{PhysRep}. 
However, this very fact will allow us
to make some drastic simplifications in the presentation 
and to concentrate 
on the points that are most essential in order to understand the work
that is to follow.

\subsection{WZW models and Toda theories}

Our starting point is a WZW model based on a real Lie 
group $G$. We have the standard action
\be 
\label{eq2}
S_{\WZW}(g) = {1 \over 2} \int d^2 x \mbox{ tr} ( \pa_+ g \pa_- g^{-1}) 
+  {1 \over 3} \int_D \mbox{tr} (g^{-1} d g )^3
\ee 
where the field $g ( x^+ , x^- ) \in G$ 
is a function of light-cone coordinates $x^\pm$ on two-dimensional
spacetime; $D$ is some
three-dimensional disc whose boundary is spacetime; and 
`tr' denotes the usual invariant inner-product on the Lie algebra 
$\cg$. 
The equations of motion 
are equivalent to the conservation conditions
$$\del_- J_+ = \del _+ J_- = 0$$
for the Kac-Moody (KM) currents defined by 
$$J_+ = k (\pa_+g )g^{-1} \, , \qquad J_- = -k g^{-1}\pa_-g \, ,$$
which take values in $\cg$.
The constant $k$ is the level.
If we introduce a
basis $\{t_a\}$ for \cg~and expand the current $J_\pm$ on this basis:
$$
J_\pm(x^\pm) = \sum_a J^a_\pm(x^\pm) t_a
$$
then the Poisson-brackets of the functions $J^a_\pm(x^\pm)$ can be given
in the form (suppressing $\pm$ indices for convenience)
$$
\{ J^a(x), J^b(y) \}_{\rm PB} = k \eta^{ab} \delta'(x-y) + 
{f^{ab}}_c J^c(x) \delta(x-y)
$$
where the Killing form 
$\eta^{ab}$ 
has been used to raise and lower indices on the structure constants
${f_{ab}}^c$ (our conventions for Lie algebras will be specified in more
detail in the next section). This is the standard form for the Kac-Moody
algebra $\widehat \cg$.
The Kac-Moody symmetry generated by these currents is 
\bea
\label{kmsym}
g(x^+ , x^-) \rightarrow a(x^+) g(x^+ , x^-) b(x^-)^{-1} 
\quad \mbox{ where  } \quad a(x^+), b(x^-) \in G \, .
\ena 
The model is also conformally invariant, with the fields $g$ behaving 
as scalars. The traceless 
energy-momentum tensor takes the well-known Sugawara form
$$T_{\pm \pm} = {1 \over 2k} \mbox{tr} (J_\pm J_\pm) \, .$$

The first step in the hamiltonian reduction of the WZW theory 
is to choose an {\em integral grading\/} 
of the Lie algebra $\cg$ by specifying a
generator $M \in \cg$ for which $\ad_M$ has integer eigenvalues. 
This defines a decomposition 
$$
\cg = \sum_n \cg_n \quad {\rm where} \quad 
X \in \cg_n \Leftrightarrow \ad_M (X) = [ M , X ] = n X \, , 
$$
so that the integers $n$ label the eigenspaces of $\ad_M$. 
It follows that 
$$
X \in \cg_m \, , \, Y \in \cg_n \quad \Rightarrow \quad  
[ X , Y ] \in \cg_{m+n} \, \mbox{ and } 
\mbox{tr} (XY) = 0 \mbox{ if } m + n \neq 0 \, .
$$
This allows $\cg$ to be a written as a direct sum of 
subalgebras of zero, strictly positive, or strictly negative grade:
$$
\cg = \cg_+ \oplus \cg_0 \oplus \cg_-  \, , \qquad \cg_+ =
\sum_{n > 0} \cg_n \, , \qquad \cg_- = \sum_{n < 0} \cg_n \, .
$$ 
We emphasize that $\cg_0$ and $\cg_\pm$ are closed under the Lie
bracket---in fact the latter two are nilpotent subalgebras---and we
can therefore associate with them subgroups $G_0$ and $G_{\pm}$ 
respectively. The restriction of the inner-product is non-degenerate
on $\cg_0$; it is completely degenerate on both $\cg_+$ and $\cg_-$
separately but defines a non-singular pairing between 
these subalgebras.

The second step is to impose constraints on the Kac-Moody currents 
of the form 
\be 
\label{cons1}
\mbox{tr} \, \left \{ X_\mp (J_\pm - k M_\pm) \right \} = 0 \, , \quad
X_\pm\in\cg_\pm \, , 
\ee 
where $M_{\pm}$ are specific elements chosen   
in $\cg_{\pm 1}$, but $X_\pm$ 
are arbitrary elements in 
$\cg_{\pm}$. 
These constraints fix exactly those parts of the currents $J_\pm$ living
in the subalgebras $\cg_\pm$ to be the constant elements 
$M_\pm$. 
The new theory defined in this way 
is invariant under a modification of the original
conformal symmetry which is constructed so as to commute with the
constraints. This is achieved by shifting the energy-momentum
tensor by a term proportional to the grading operator, so that it
becomes $$T_{\pm \pm} = {1 \over 2k} \mbox{tr} (J_\pm J_\pm) - \mbox{tr}
(M \del_\pm J_\pm )$$
The constraints above are first-class, corresponding to
a gauged KM symmetry for the nilpotent 
subgroups $\cg_\pm$. 
In what follows we shall always assume in addition 
the {\em non-degeneracy condition}
\be
\label{nondg}
\cg_\mp \cap \ker(\ad_{M_\pm}) = \{ 0 \}
\ee
The importance of this can be explained in a number of different ways,
but in particular it guarantees that the residual KM symmetry 
in the presence of the constraints becomes a $\cw$-algebra with a basis
of primary fields.
At a technical level, this is revealed through the 
fact that the non-degeneracy condition allows us to impose
{\em highest weight} or 
{\em Drinfeld-Sokolov} (DS) gauges, which reveal a direct 
connection between the KM and $\cw$-algebra generators,
as we shall indicate below.

Let us now recall how to make manifest the gauge 
symmetry corresponding to the 
constraints (\ref{cons1}).
We introduce gauge fields $A_\pm$ taking values in
$\cg_\pm$ respectively and then consider the action 
\bea
\label{gaugewzw}
S(g,A_+,A_-) = S_{\WZW}(g) 
\phantom{XXXXXXXXXXXXXXXXXXXXXXXXXXX} \nn
+ \int \! d^2x \mbox{ tr} \left \{ \,  
A_- ( M_+ - (\pa_+ g ) g^{-1} ) + 
A_+ ( M_- + g^{-1} \pa_-g ) 
+ A_-gA_+g^{-1}\right \} 
\ena
which we claim is equivalent to the constrained WZW model. 
The gauged action above is invariant under the 
local transformations 
\bea
\label{gaugesym}
g \rightarrow u g v^{-1} , \quad
A_+ \rightarrow v A_+ v^{-1} - (\del_+ v) v^{-1} , \quad
A_- \rightarrow u A_- u^{-1} - (\del_- u) u^{-1} , \nn
\mbox{ where } \quad 
u (x^+ , x^-)\in G_-  , \,\, v (x^+ , x^-) \in G_+ \, . 
\ena 
Although this is easy to check, it relies on a number of unusual 
features:
the light-cone components $A_{\pm}$ are defined in different
subalgebras of the full Lie algebra $\cg$; the terms involving
$M_{\pm}$ are put in by hand, being 
gauge-invariant by themselves; and invariance
of these and other terms in the lagrangian depends crucially on
properties of the grading. 
Now any single component of a gauge field can always be
set to zero (locally) by a gauge transformation and since $A_\pm$ 
live, by definition, in disjoint subalgebras, 
we are always free to make the gauge choice 
$A_\pm = 0$. On doing so, we recover the previous action $S_{\WZW}$, but in
addition we must impose the equations
of motion for $A_\pm$, which in this gauge give exactly the
desired constraints (\ref{cons1}).

We have thus established that the constrained WZW model is 
equivalent to 
the gauged lagrangian (\ref{gaugewzw}) on making a
specific gauge choice. 
But there is another natural gauge choice which leads directly to
the description in terms of a non-abelian Toda theory.
Given the grading of $\cg$ introduced earlier, we
can define (locally) a Gauss decomposition of an arbitrary group
element: 
$$
g = a\,g_0\,b, \quad \mbox{where} \quad 
a\in G_- , \,\,g_0\in G_0, \, \,b\in G_+.
$$
Now we can clearly use the gauge freedom (\ref{gaugesym}) to set $a = b = 1$.
Unlike the previous gauge choice, the fields $A_\pm$ no longer vanish, but
they can be eliminated algebraically from the lagrangian. 
The result is a (non-abelian) Toda theory for 
the field $g_0 \in G_0$
governed by the action
\be
\label{eq1}
S(g_0) = 
S_{\WZW}(g_0) - \int d^2x \tr \{ M_+  g_0 M_-g_0^{-1} \} .
\ee
The integrability of these models was first discussed in \cite{LS} 
from a rather different point of view.

We emphasize that the first term in the Toda action is again of WZW
type, though now for the zero-grade subgroup $G_0$. 
This part of the action has a KM symmetry $\widehat \cg_0$, but 
the potential term involving $M_\pm$ 
breaks this down to a smaller symmetry.
If $G_0$ is non-Abelian, the theory as a whole will in general have a 
residual Kac-Moody invariance $\widehat \ck$ contained within its 
chiral symmetry algebra, 
where the subalgebra $\ck \subset \cg_0$ is the centralizer of $M_\pm$.

\subsection{$sl(2)$ embeddings} 

The data specifying the grading and constraints 
on the original WZW model 
can be expressed in terms of an 
embedding of $sl(2,\fl{R})$ into \cg.
We consider a set of generators  
$\{ M_0, M_+, M_- \}$ which obey 
$$
[M_0 , M_\pm ] = \pm M_\pm \ , \qquad [M_+ , M_- ] = 2M_0 \ , 
$$ 
and which are therefore a basis for an $sl (2, \fl{R})$ subalgebra 
$\cs \subset \cg$.
As the notation suggests, $M_\pm$ should be identified 
with the specific Lie algebra elements used earlier in our
definition of the constraints (\ref{cons1}). 
We can decompose \cg~into irreducible representations of 
\cs , with Ad${}_{M_0}$ having integer or half-integer eigenvalues,
and we then say that 
$\cs \subset \cg$ is an {\em integral} or {\em half-integral} embedding
respectively.
Given an integral embedding, we can immediately define an integral grading 
$M=M_0$, which always satisfies the non-degeneracy condition (\ref{nondg}).
This is the class of reduced WZW models we shall focus on in this
paper.
We should emphasize that we will always understand the 
embedded subalgebra $\cs$ to have a preferred set of generators as
above,
so that we are really specifying not just the subalgebra but also the
particular element $M_0$ which defines the grading.

There is actually a more general possibility for the choice of grading
whenever there exists a generator 
$Y \in \cg$ that commutes with the entire subalgebra \cs. 
Then we can also consider an integral grading operator of the
form $ M = M_0 + Y $, provided (\ref{nondg}) still holds.
One can show \cite{U1} that taking a non-zero $Y$ does not change 
the \cw-algebra in the reduced theory, so from this point of view 
one gains nothing new.
In certain cases, however, the addition of $Y$ can be used to define 
an integral grading from a 
half-integral embedding, giving a more general class of lagrangian
realizations. 
It can also be shown that to any choice of integer
grading $M$ and constraints $M_\pm$ which together satisfy (\ref{nondg}),
one can always associate an  
embedded subalgebra \cs~in the manner described above (see e.g.
sections 3.3, 3.4 and appendix B of 
\cite{PhysRep} for more details). 
Notwithstanding these comments, it is natural to restrict
attention in the first instance to the case $Y=0$. 
In fact it is also true that $Y\neq 0$ leads to no new 
possibilities as far as the classification given in
section 5 is concerned.

{}From what we have said so far, we are confronted with the task of 
understanding the possible $sl (2, \fl{R})$ subalgebras $\cs$ in $\cg$.
Embeddings of $sl (2, \fl{C})$ in complex Lie
algebras have been classified by Dynkin 
\cite{Dy}, and one may extend his results immediately  
to the case where 
$\cg$ is real but maximally non-compact. 
The classification states that any embedding of $sl(2)$ into 
\cg~can be realized as a {\em principal} embedding in a 
{\em regular} subalgebra $\ch \subset \cg$, up to a few exceptions
which can be enumerated\footnote{The exceptions are ${n-2\over 2}$
embeddings in $D_n$, one embedding in $E_6$, and two embeddings in
each of $E_7$ and $E_8$, which are not principal in any regular
subalgebra \cite{Dy}.}. 
We recall that a regular subalgebra is one for which 
$\DE({\cal H}) \subset \DE({\cal G})$, where $\DE$ denotes 
the set of all roots, while a principal 
$sl(2)$ embedding is one in which $M_{\pm}$ are taken to be linear
combinations of step operators for all the positive/negative simple
roots in some chosen basis. 
We shall denote an embedding of this important kind simply 
by the pair (\cg,\ch).
Abelian Toda models correspond to
reductions of the
form (\cg,\cg), and are so called because the group $G_0$ is in this
case just a maximal torus in $G$. 
In section 5 we shall classify the integral embeddings 
$(\cg , \ch)$ with $\cg$ classical and $\ch$ simple.
The general results given in this section, however, will 
be independent of any such more detailed 
information about an embedding $\cs \subset \cg$.

\subsection{\cw-algebras and conformal field theory} 
\label{tjm}
We have seen above that to every integral embedding $\cs \subset \cg$
we can associate a non-abelian Toda model.
We now recall how this leads to a realization of an extended
conformal algebra which we denote by $\wgs$.
We need to clarify how the generators of the \cw-algebra appear,
and how this relates to the usual approach in CFT.

Up until now we have been working in Minkowski
spacetime, with Kac-Moody currents $J_\pm (x^\pm)$ taking 
values in the the real Lie algebra $\cg$. If we introduce a 
basis of generators $t_a$ for $\cg$ then we can 
write $J_\pm = \sum_a J_\pm^a t_a$ and the components $J_\pm^a$ are
real classical fields, or hermitian operators upon quantization. 
To make contact with the usual language of conformal
field theory, we take our model to be defined on a circle in the
spatial direction, with all fields being periodic. After complexifying
the coordinates $(x^\pm) \rightarrow (w , \bar w)$ we use 
a conformal transformation to map from the cylinder to the plane 
$(w , \bar w) \rightarrow (z ,\bar z) = ( e^{i w} , e^{- i \bar w})$.
Now our KM currents are the familiar holomorphic and anti-holomorphic 
quantities 
$$J (z) = J^a (z) t_a  , \qquad \bar J ( \bar z) = \bar J^a
(\bar z) t_a$$
and for our purposes we may set the level $k=1$ without loss of 
generality.
The energy-momentum tensor similarly has components $T(z)$ and 
$\bar T( \bar z)$. 

An important point which deserves emphasis is that 
all fields regarded as functions on the complexified plane 
$(z, \bar z)$ 
have a natural notion of hermitian conjugation which is 
inherited from their original definition as fields on Minkowski spacetime.
For a holomorphic field $\Phi$ which is
quasi-primary with conformal
weight $h$, the hermitian conjugate $\Phi^*$ is defined by
\be
\label{conj}
\Phi^* (z) = (i/z)^{2h} \Phi (1/z) \ .
\ee
We stress that $z$ and 
$\bar z$ are independent complex variables and that the
conjugation operation on fields is defined entirely within each
holomorphic or anti-holomorphic sector. Real fields in Minkowski space
become self-conjugate, satisfying $\Phi = \Phi^*$. In 
particular this applies to the Kac-Moody
currents $J^a (z)$ with $h = 1$. It also applies to the
energy-momentum tensor $T(z)$ with $h=2$,
giving the standard hermiticity properties for the Virasoro
generators.

One very powerful aspect of the conformal reduction of WZW models 
is that much of the structure of \cw-algebras which emerge 
can be understood simply by analyzing how
the adjoint representation of \cg~decomposes into irreducible 
representations of the embedded subalgebra \cs .
We shall therefore choose a basis of generators $\{t_a\}$ 
to be states of definite total spin $j$ and $M_0$ eigenvalue $m$ 
with respect to $\cs$. In addition, we need a label $\alpha$ to 
distinguish representations of the same spin.
Our generators will now be written $\{t^{\A}_{(j,m)}\}$.
The subalgebra $\cs$ itself is of course a
representation of spin $j = 1$ which we label by $\A=0$, 
so that $t^{0}_{(1,0)} = M_0$ and $t^{0}_{(1,\pm 1)} = M_\pm$.
Concentrating on the holomorphic sector,
the Kac-Moody current can be expanded 
$J(z) = J^{\A}_{(j,m)}(z) t^{\A}_{(j,m)}$ with respect to this basis,
and the constraints 
(\ref{cons1}) then take the form: 
\be
\label{reps}
J^{\A}_{(j,m)}(z) = \left \{ \begin{array}{l}
1 \mbox{ if } \A = 0 \ \ (j,m) = (1,-1) \\ 
0 \mbox{ else } \end{array} \right. \quad \quad \mbox{for } m < 0.
\ee

As mentioned before, these constraints are first-class,
corresponding to a gauge symmetry of the constrained action.
The KM currents themselves transform under this gauge symmetry,
reflecting its non-abelian nature.
The generators of the symmetry algebra $\wgs$ appearing in the 
constrained model 
can be defined as the set of gauge-invariant 
differential polynomials in the KM currents.
It is more enlightening to make this concrete, however,
by introducing some specific way of fixing the gauge symmetry.
Any particular gauge choice allows the gauge-invariant polynomials
to be identified with certain components of the KM currents,
and the classical $\cw$-algebra is then given by the Dirac brackets
of these quantities, defined using the full set of constraints.

A very convenient gauge choice in which the generators have a
particularly simple structure is 
the {\em highest-weight} Drinfeld-Sokolov
gauge, where 
the only non-zero current components correspond to generators of 
highest weight with respect to \cs, i.e. 
\be
\label{dshw}
J_{hw}(z) = M_- + \sum J^{\A}_{(j,j)}(z) \, t^{\A}_{(j,j)}. 
\ee
One can show that there is exactly one independent 
gauge-invariant polynomial for each of the highest weight components, 
so that the independent generators of $\wgs$ are simply
$W^{\A}_{j+1}(z) = J^{\A}_{(j,j)}(z)$.
Moreover, it is easy to show that 
the fields $W^\A_h$ have (holomorphic) conformal weight $h = j+1$,
so that the spin-content of the \cw-algebra 
can be simply
read off from the representation theory associated with 
the embedding $\cs \subset \cg$, see \cite{FRS}.

Two examples are worth mentioning. 
First, the energy-momentum tensor
$T$ of the 
extended conformal algebra appears as the gauge-invariant polynomial 
corresponding to the highest weight generator $M_+$---i.e.~to the
representation carried by $\cs$ itself, which we have labelled 
$\alpha = 0$. 
In this case $j = 1$ (the adjoint representation) and of course $T$
has $h = 2$ (although it is only quasi-primary). 
Second, consider the centralizer $\ck \subset \cg_0$ of \cs~in \cg. 
Since it commutes with \cs, its
generators must appear as singlets, with $j = 0$, in the highest weight 
decomposition. Consequently, these generators have conformal weight 
$h = 1$ and we expect them to be KM currents. 
This corresponds exactly to the 
residual $\widehat \ck$ Kac-Moody symmetry 
of the non-abelian Toda action to which we drew attention earlier. 

It is important to stress that the simple relation between highest-weight 
generators and gauge-invariant polynomials holds only in the 
highest weight gauge. 
In a general DS gauge, 
the gauge-invariant polynomials take the form:
\be
\label{Whw}
W^{\A}_{j+1}(z) = J^{\A}_{(j,j)}(z) + \ldots,
\ee
where the dots denote terms which are non-zero in general, but which
are higher order in currents 
and/or derivatives. It has been shown \cite{BoTj} that this form of the 
generators holds also in the quantized algebra. 
This gives the generators of the \cw-algebra as differential 
polynomials in all the currents in $\widehat{\cg}$. 

Apart from the DS class of gauges, another important choice 
is the diagonal gauge, where only currents 
corresponding to generators with zero grade are non-vanishing. 
In this case we can write the current as: 
$$
J_{diag}(z) = M_- + \sum J^{\A}_{(j,0)}(z) \, t^{\A}_{(j,0)}. 
$$
Using this gauge it is possible to see that
if we restrict the general differential polynomial expressions 
(\ref{Whw}) 
involving the $\widehat \cg$ currents to just those with zero grade, 
i.e.~we keep only the currents from $\widehat{\cg}_0$, then this gives an
equivalent realization of \cw. 
This restricted realization is clearly the more natural 
point-of-view when considering
the non-abelian Toda action, with fields in $G_0$, 
rather than the original gauged WZW theory based on $G$.
The equivalence between these two different sets of variables 
is essentially the so-called generalized Miura transformation. 

Finally, we recall that
the structure of the original Lie algebra $\cg$
is always present in algebra $\wgs$ in the following way.
Given an extended conformal algebra,
we can define the (linearized) {\em vacuum preserving algebra\/} (VPA) 
by expanding each generator $W^{\A} (z)$
of conformal weight $h$ in modes: $W^{\A} (z) = \sum_n W^{\A}_n 
z^{-n-h}$ and considering the finite-dimensional subalgebra which is generated
by $W^{\A}_n$ with $|n| \leq (h-1)$ on dropping all quadratic and
higher-order terms. This vacuum-preserving algebra is 
isomorphic to the original Lie algebra $\cg$, as explained in \cite{FeOrIt}. 
Furthermore, the 
VPA always contains the subalgebra $\cs$, whose generators $\{ M_0 ,
M_\pm\}$ correspond precisely
to the usual $sl(2)$ subalgebra of the Virasoro algebra with
generators $\{ L_0 , L_{\pm 1} \}$.
These remarks will prove important later.

\section{Real Forms} 
\label{sec-2}
\subsection{Real Forms of Lie algebras and $sl(2)$ embeddings}
\label{sec-2-1}

We first recall some basic facts concerning the relationship between 
a real Lie algebra $\cg$ and its {\em complexification}, the complex Lie
algebra $\cg^{\bf c} = \cg \otimes \fl{C}$ 
(for more details see e.g.~\cite{Hel,SW}). We then give methods 
for constructing real forms which we shall later use extensively. 

For any real Lie algebra $\cg$ or complex Lie algebra $\cg^{\bf c}$, the 
Killing form will be denoted by $\eta( X , Y ) = \lambda \, \mbox{tr} (XY)$ 
with $\lambda$ a suitable constant. 
As a basis for either of these algebras we introduce 
an orthonormal set of generators $\{ t_a \}$ obeying 
\bea
[ t_a , t_b ] = f_{ab}{}^c t_c \quad &{\rm where}& \quad f_{ab}{}^c
\in {\fl R} 
\label{struct} \\
\eta(t_a,t_b) = \eta_{ab} = \sigma_a \D_{ab} \quad & {\rm where}& 
\quad \sigma_a = \pm 1
\label{kill} 
\ena
If $d_\pm$ denotes the number of basis elements for which 
$\sigma_a = \pm 1$ 
then the real dimension of the real Lie algebra
$\cg$ is $\mbox{dim} (\cg) = d_++d_-$.
The {\em character\/} of 
$\cg$ is defined to be the signature of the Killing form, 
$\sigma (\cg) = \sum_a \sigma_a = d_+-d_-$. 
The character provides a way of distinguishing between 
different real forms and hence of classifying them.

We shall reserve the symbols $A_n$, $B_n$, $C_n$, $D_n$ 
for complex Lie algebras, while their real forms will be denoted by
the symbols given below together with their 
characters. 
\be
\label{tab0} 
\begin{array}{|c|l|c|}
\hline 
\cg^{\bf c} & \cg  & \sigma  \\ \hline\hline
A_n & sl(n\!+\!1) \equiv sl(n{+}1,\fl{R}) & n \\ \cline{2-3} 
&su(p,q)\,\, \, p\!+\!q = n\!+\!1 & 1-(p - q)^2 \\
\cline{2-3}
&su^*(n{+}1)\, \, \, n \mbox{ odd }& -n-2 \\ \hline
B_n & so(p,q) \, \, \, p\!+\!q = 2n\!+\! 1
& n + \frac{1}{2} - \frac{1}{2}(p-q)^2 \\ \hline
C_n & sr(n) \equiv sp(2n, \fl{R}) & n \\ \cline{2-3}
    & sp(p,q) \, \, \, p\!+\!q = n 
& - 2 [ n + (p-q)^2 ] 
\\ \hline
D_n & so(p , q )\, \, \, p\!+\!q = 2n 
& n - \frac{1}{2}(p-q)^2 \\ \cline{2-3}
& so^*(2n) & -n \\ \hline
\end{array}
\ee
\noindent
It is clear from the table that the character $\sigma$ usually
specifies the real form $\cg$ uniquely given the complex type
$\cg^{\bf c}$.
The only exceptions occur in the 
$A_n$ and $D_n$ series. For $n = 4k^2{-}3$, the algebras 
$su^*(n{+}1)$ and $su(k^2\!+\!k\!-\!1,k^2\!-\!k\!-\!1)$ 
have the same character but are not isomorphic for $k > 1$.
For $n=k^2$ the algebras $so^*(2n)$ and $so(n\!+\!k,n\!-\!k)$ also 
have the same character without being isomorphic.
In these cases the algebras can be distinguished by their maximal 
compact subalgebras.

We shall use the symbol $U_1$ to mean a complex Lie
algebra with a single generator. This has only one distinct real form
in any intrinsic sense, of course, but it is nevertheless useful 
to distinguish a `compact' real form 
$u(1)$ and a `non-compact' form $gl(1)$ which arise
through an embedding into a non-abelian algebra whose Killing form is
negative-definite or positive-definite respectively when restricted to
these abelian subalgebras.

A convenient way to define a real form of a given complex 
Lie algebra $\cg^{\bf c}$ is via a {\em conjugation map}, by which we
mean an anti-linear automorphism $\tau$ of order two on $\cg^{\bf c}$:
\beano 
\label{conjmap}
[\tau(X),\tau(Y)] & = & \tau([X,Y]), \quad\quad X,Y \in \cg^{\bf c}.\\
\tau(aX+bY) & = & a^* \tau(X) + b^* \tau(Y), \quad\quad a,b \in \fl{C}. \\
\tau^2 & = & 1
\enano
An automorphism of this type essentially corresponds to a notion of complex
conjugation on $\cg^{\bf c}$, which explains its name 
(we follow \cite{SW} in this). 
Given a conjugation map, 
we can define a corresponding real Lie algebra $\cg^\tau$ consisting
of the elements it leaves invariant:
\be 
\cg^\tau = \{ \, X\in\cg^{\bf c} \, | \, \tau(X)=X \, \} \ .
\label{fixpt}\ee 
The statements that $\cg^\tau$ is a real subalgebra of $\cg^{\bf c}$ 
and that $\tau$ is an anti-linear automorphism
of $\cg^{\bf c}$ are equivalent, since $x, y \in \cg$ means that
$[x, y] = [ \tau(x), \tau(y)] = \tau ([x,y]) \in \cg $.
It is also easy to see that any real form of $\cg^{\bf c}$ 
arises in this way \cite{Hel,SW}. 
(We shall often omit the label 
$\tau$ and write the resulting real Lie algebra simply as $\cg$ 
when no possible confusion can arise.) 

A conjugation map 
$\tau$ can be defined by its action on some 
basis $\{ t_a \}$ for $\cg^{\bf c}$ 
as introduced in (\ref{struct}) and (\ref{kill}) above, 
the extension to the whole 
algebra being fixed by the property of anti-linearity. We write
\be 
\tau(t_a) = \sum_b t_b \tau_{ba}. 
\ee
where $\tau_{ab}$ is an $(n{\times}n)$ matrix which must satisfy
$\tau_{ab}^* \tau_{bc} = \delta_{ac}$.
A general element of the corresponding 
real Lie algebra $\cg^\tau$ is then 
\be
\label{real}
\Phi = \sum_a \phi_a t_a, \quad \mbox{ where } \quad
(\phi_a)^*= \sum_b \tau_{ab}\phi_b. 
\ee 
It is sufficient for our purposes to consider
a basis $\{t_a\}$ for which the  
Killing form and the conjugation map are simultaneously diagonalized,
so that in addition to (\ref{kill}) we have:
\be\label{conjbasis} 
\tau_{ab}=\epsilon_a\D_{ab}, \qquad {\rm where} \qquad \eps_a=\pm 1
\ee 
Notice that $\tau (t_a) = t_a$ automatically 
extends to an anti-linear automorphism because the structure
constants $f_{ab}{}^c$ are real. 
The resulting real form $\cg$ is simply the real span of the basis $\{
t_a \}$ with character $\sigma (\cg) = \sum_a \sigma_a$ as discussed in the
first paragraph. 
More generally, however, we see from (\ref{conjbasis}) that the real form
$\cg^\tau$ has character 
\be
\label{char}
\sigma (\cg^\tau) = \sum_a \sigma_a \epsilon_a. 
\ee

Of particular importance for any semi-simple complex 
Lie algebra $\cg^{\bf c}$ is the Cartan-Weyl basis. 
Let $H$ be a choice of Cartan subalgebra and 
$\Delta$ the corresponding set of non-vanishing roots.
For each $\alpha \in \Delta$ we have a Cartan generator
$h_\alpha \in H$ and step operators $e_{\pm \alpha}$ obeying
\be 
[h_\alpha , e_{\beta} ] = {(\alpha \cdot \beta)}  
e_{\beta} \, , \quad
[e_\alpha , e_{-\alpha} ] = h_\alpha
\, , \quad
[e_\alpha , e_\beta ] = N_{\alpha \beta} \, e_{\alpha + \beta} 
\, \, \, (\alpha + \beta \neq 0)  
\ee
where $N_{\alpha \beta}$ is non-vanishing iff $\alpha + \beta \in
\Delta$. 
Let $\{ \alpha_j \}$ be a set of simple roots with $j = 1,\dots, 
\mbox{rank} (\cg^{\bf c})$; it is convenient to 
set $h_j = h_{\alpha_j}$. For each root we have 
\be 
\alpha = \sum_j \ell_j (\alpha)  \alpha_j \, ,
\qquad
h_\alpha = \sum_j \ell_j (\alpha) h_j \, ,
\qquad
\ell (\alpha) = \sum_j \ell_j (\alpha) 
\ee 
where $\ell_j (\A)$ are integers and $\ell(\alpha)$ is the height of
the root. 
The subset of positive roots ($\ell(\alpha) \geq 0$) 
will be denoted $\Delta_+$. 
With this normalization for the generators the 
non-trivial entries of the Killing form are 
$\eta( h_i , h_j ) = (\alpha_i \cdot \alpha_j)$, 
a positive-definite matrix, and
$\eta(e_\A,e_{-\A}) = 1$ for each root
$\alpha$.
It follows that we can write an orthogonal 
decomposition of the space of generators
$
\{ h_j \} \oplus \{ e_\A +e_{-\A} \} \oplus \{ e_\A - e_{-\A}\}
$
and by taking appropriate real combinations within each of these
subspaces the Killing form can be diagonalized 
with eigenvalues $+1$, $+1$, $-1$ respectively.

There are two real forms which arise in a
uniform way for any complex Lie algebra $\cg^{\bf c}$ by using the
Cartan-Weyl basis. 
Consider first a conjugation map
which fixes each of the Cartan-Weyl generators: 
\be \label{split} 
\tau(h_j) = h_j, \qquad \tau(e_{\pm \A}) = e_{\pm \A} 
\ee
This defines a real Lie algebra consisting of the real span of 
the Cartan-Weyl generators. 
The character is 
$\sigma = \mbox{rank}(\cg^{\bf c})$; this is the {\em maximally 
non-compact} or {\em split} real form. 
Alternatively, if we consider the conjugation map 
\be \label{comp}
\tau(h_j) = - h_j, \qquad \tau(e_{\pm \A}) = - e_{\mp \A} \quad
\Leftrightarrow \quad 
\tau(e_\A\pm e_{-\A}) = \mp (e_\A \pm e_{-\A})
\ee
then the resulting real Lie algebra is the real span of 
$
\{ ih_j , i(e_\A+e_{-\A}) , (e_\A-e_{-\A}) \}
$.
{}From the remarks above it is clear that 
in this case the Killing form 
is negative-definite and the character is 
$\sigma = -\mbox{dim}(\cg^{\bf c})$; this 
is the {\em compact} real form. 

In each of these two cases the conjugation map 
was chosen to fix the Cartan subalgebra $H$. 
In fact there is no loss of generality in assuming this,
since any automorphism is conjugate to one with this property.
Furthermore, the fact that $\tau (H) = H$ means that the algebra 
automorphism specifies a {\em root automorphism}, by which we mean 
an element of Aut($\DE$), the group of
linear transformations which fix the set of roots $\DE$.
This root automorphism will, with 
a slight abuse of notation, also be denoted by $\tau$.
The root automorphism does not quite specify the algebra 
automorphism uniquely, however; the most general possibility is
\be
\label{Waut}
\tau(h_{\A_j}) = h_{\tau(\A_j)},\quad \tau(e_\A) = \chi_\A e_{\tau(\A)}, 
\quad \tau(e_{-\A}) = \chi_\A e_{\tau(-\A)} \quad {\rm where} \quad 
\chi_\alpha = \pm 1 \ .
\ee
The constants $\chi_\alpha$ must satisfy the consistency condition
\be
\label{def-chi}
\chi_{\A+\B} = \frac{N_{\tau(\A)\tau(\B)}}{N_{\A\B}} \chi_\A\chi_\B
\ee
but we may choose $\chi_{\A_i}$ independently for each simple root, and then 
the remaining values are determined.
Finally, the algebra automorphism $\tau$ 
defined in this way need not necessarily be of order two, even if the 
root automorphism in $\mbox{Aut}(\DE)$ is. 
To ensure this one must have in addition 
\be
\label{cons}
\chi_\A = \chi_{\tau(\A)}
\ee 
for all $\A$. 
It is not difficult to see that it is 
sufficient to impose this condition for each of the simple
roots. 

We should now draw attention to the relationship 
between Aut$(\DE)$, defined above, and the Weyl group 
$W(\DE)$, which is generated by reflections in the (hyper)planes 
perpendicular to each root.
The Weyl group $W(\DE)$ is a normal subgroup of Aut$(\DE)$ and the
quotient $\mbox{Aut}(\DE)/W(\DE)$ can be identified with
outer-automorphisms of the algebra, 
or more pictorially with the symmetry group of its 
Dynkin diagram. It is therefore non-trivial only for the (simple) algebras
$A_n$, $D_n$ and $E_6$.
We shall refer to any representative of this quotient
group as a {\em diagram automorphism}, for obvious reasons.

We can now set the split and compact real forms in a more general context.
The split real form corresponds to the root automorphism
$\tau (\alpha) = \alpha$, which is simply the identity in the Weyl group.
It is clearly consistent to choose $\chi_\alpha = 1$ as in 
(\ref{split}).
The compact real form corresponds to the root automorphism
$\tau(\alpha) = -\alpha$. For this element of $\mbox{Aut}(\DE)$, 
the general solution of the
consistency condition (\ref{def-chi}) is 
\be
\label{chi} 
\chi_\A = (-1)^{ \ell(\A)+1} \prod_j (\chi_{\A_j})^{\ell_j (\A)} 
\ee
The compact form corresponds to the particular choice
$\chi_\A = -1$ for all roots, 
but different choices of the $\chi_\alpha$ could lead to 
other conjugation maps and real forms.

We are now in a position to calculate the character 
of a real form defined by any conjugation map of the general
kind (\ref{Waut}). 

\begin{theo}
Let $\tau$ be a conjugation map on $\cg^{\bf c}$ which fixes a Cartan
subalgebra $H$ and which is defined from a root automorphism 
by the quantities $\chi_\alpha$ as in {\rm (\ref{Waut})}.
The character of the corresponding real Lie algebra $\cg$ is
\be
\label{char2}
\sigma = {\rm tr} (\tau_H) + 2 \!\sum_{\A\in\Gamma_-}\chi_\A 
\ee
where $\tau_H$ is the restriction of $\tau$ to $H$ and 
$\Gamma_- = \{ \A \in \Delta_+ | \tau(\A) = -\A\}$. 
\end{theo}

\noindent
Proof: The key idea is 
to calculate the contributions to the character from combinations 
of root generators which are diagonal with respect to both $\tau$ and $\eta$.
If $\tau (\alpha) = \alpha$ and 
$\tau(e_\A) = \chi_\A e_\A$ then the combinations 
$(e_\A \pm e_{-\A})$ 
have equal eigenvalues $\chi_\A$
under $\tau$ but opposite eigenvalues $\pm 1$ 
for the Killing form. As a result we see from 
(\ref{char}) that their 
contributions to the character always cancel.
If $\tau(\A) = -\A$, on the other hand, then 
$\tau(e_{\pm \alpha}) = -\chi_\A e_{\mp \A}$ and now the 
combinations 
$(e_\A \pm e_{-\A})$ 
have opposite eigenvalues $\pm \chi_\alpha$ under $\tau$.
The total contribution to the character from
these generators is therefore $2\chi_\A$ for each $\A \in \DE_+$. 
Finally we consider those roots for which $\tau(\A) \not = \pm \A$. 
In this case there are four linearly-independent combinations 
$\hf(e_\A + \nu_1 e_{\tau(\A)} + \nu_2 e_{-\A} 
+ \nu_1\nu_2 e_{-\tau(\A)})$ with $\nu_i=\pm 1$.
The eigenvalue of $\tau$ is $\nu_1\chi_\A$ 
while the eigenvalue of $\eta$ is $\nu_2$. The
contributions to the character from all four combinations therefore cancel. 
The result now follows. \hfill $\Box$\\[2pt]

Note that the formula in Theorem 1 successfully reproduces
the two cases we encountered earlier.
If the root automorphism $\tau$ is the identity in the 
Weyl group then $\Gamma_-$ is empty and $\sigma = \mbox{rank}
(\cg^{\bf c})$ as required for the split form.
If we take instead the root automorphism 
$\tau(\A)=-\A$ then for the general case given by
(\ref{chi}) we find
that eq.~(\ref{char2}) becomes 
\be
\label{chi2} 
\sigma = -\mbox{rank}(\cg) + 2 \! \sum_{\A\in\DE_+} (-1)^{\ell(\A)+1} 
\prod_j (\chi_{\A_j})^{\ell_j (\A)} 
\ee
For the choice $\chi_\A = -1$ we find 
$\sigma = - \mbox{dim}(\cg)$, as required for the compact real form.

Having developed some technology to discuss the real forms of Lie
algebras, we recall that our chief concern in this paper 
is with the integrable models and
$\cw$-algebras which are associated to 
embeddings of $sl(2,\fl{R})$ into real Lie algebras, 
with a preferred basis $\{M_0 , M_+ , M_- \}$ which selects 
the grading operator $M_0$.
There is a very simple idea which we shall use later 
to obtain from the well-known classification
of complex embeddings a classification of (a large subset
of) real embeddings.
An embedding $\cs \subset \cg$ extends automatically
to an embedding of complex algebras 
$\cs^{\bf c} \subset \cg^{\bf c}$, and 
the preferred basis 
$\{M_0 , M_+ , M_- \}$ 
for the real algebra $\cs \cong sl(2,\fl{R})$ becomes a 
preferred basis for the complex 
algebra $\cs^{\bf c} \cong sl(2, \fl{C})$.
Since much more is known about complex embeddings, we would like
to reverse this procedure, 
i.e., to take an embedding 
$\cs^{\bf c} \subset \cg^{\bf c}$ with a preferred basis
and obtain from it an embedding of real forms $\cs \subset \cg$.
It is clear that what we require is the following. 

\begin{defi}
An embedding $\cs^{\bf c} \subset \cg^{\bf c}$ is compatible
with a real form $\cg$ or the conjugation map $\tau$ which defines it 
iff $\tau$ fixes the preferred $sl(2)$ generators:
\be
\label{sinv}
\tau (M_0) = M_0 , \quad \tau (M_\pm) = M_\pm
\ee
\end{defi}
\noindent
When this condition is met we clearly have an embedding
$\cs \subset \cg$ of the type we are seeking.

We remark that the idea of a specific grading,
as defined by $M_0$ above, arises naturally in the work of Dynkin 
on embeddings $\cs^{\bf c} \subset \cg^{\bf c}$.
We shall make use later 
of a particular result which underlies 
Dynkin's classification, which states that for any integral 
complex embedding it is
possible to choose a Cartan subalgebra of grade zero elements,
and a basis of step operators for simple roots which have gradings
0 or $\pm 1$\footnote{We repeat that we deal only with integral
embeddings in this paper; half-integral embeddings would allow
half-integral gradings of some step operators. Note also that in
Dynkin's original work the definition of the grading is double that
used here, so what we call integral gradings would correspond to even
integral gradings in Dynkin's terminology.}.

In the special case of a principal embedding
$\cs^{\bf c} \subset \cg^{\bf c}$, all the raising and
lowering operators corresponding to the simple roots have grades 
+1 and -1 respectively. This means that $M_\pm$ are linear
combinations of these step operators with non-zero coefficients.
It is easy to see that for such an embedding the only compatible 
conjugation maps correspond to diagram automorphisms.
We shall refer to the resulting real embedding $\cs \subset \cg$ 
as being principal, irrespective of whether the diagram automorphism
involved is trivial or not. 
Nearly all complex embeddings are actually of the type 
$(\cg^{\bf c}, \ch^{\bf c})$, 
meaning that $\cs^{\bf c} \subset \ch^{\bf c}$ is a principal
embedding in the regular subalgebra $\ch^{\bf c} \subset \cg^{\bf c}$.
By extension of what we have already said, any compatible real form
$(\cg,\ch)$ corresponds to a conjugation map which must act as a 
diagram automorphism on $\ch^{\bf c}$.
These observations will prove important in section 5.

Before moving on to Toda theories proper, we also 
note the following Corollary of Theorem 1.

\begin{cor}
\label{char3}
Consider a real embedding
$\cs\subset\cg$ defined by a compatible conjugation map
from a complex embedding $\cs^{\bf c}\subset \cg^{\bf c}$.
Let $\cg_0$ be the real form of the zero-grade
subalgebra $\cg_0^{\bf c}$. The characters are related by
$$
\sigma (\cg) =\sigma (\cg_0)
$$
\end{cor}

\noindent Proof: 
The automorphism $\tau$ respects the gradation, i.e. the grade of 
$\tau(e_\A)$ is equal to the grade of $e_\A$. Furthermore, the grade 
of $e_{-\A}$ is minus the grade of $e_\A$. This implies that if 
$\tau(e_\A)=\pm e_{-\A}$, then $e_\A \in \cg_0$. Since furthermore 
the Cartan subalgebra of $\cg$ is contained in $\cg_0$, the result follows 
immediately from eq. (\ref{char2}). \hfill $\Box$ 

\subsection{Real Forms and Toda Theories} 

The WZW models 
which are most obviously relevant for physical applications are those 
defined on a real compact group manifold, since these will have
positive-definite actions and the corresponding KM algebras will have
highest-weight, unitary representations.
Nevertheless, WZW 
models for non-compact groups have attracted some attention in the
past, for example as models of string-theory in curved space-time, 
see e.g. \cite{Ba}, or as a starting point for the construction 
of solvable black hole models, see e.g. \cite{BH}. 

Consider a WZW model based on a group $G$ whose real Lie algebra $\cg$
can be defined from the
complex Lie algebra $\cg^{\bf c}$ by a conjugation map 
$\tau$. 
The group elements appearing in the action can be parametrized 
locally (in a neighbourhood of the identity) by 
$g=\exp(\sum_a \phi_a(x)t_a)$, where 
the fields 
$\phi_a(x)$ satisfy the reality-conditions (\ref{real}). 
The currents of this theory are also elements of the Lie algebra,
and so they satisfy analogous reality conditions:
\be\label{kmreal} 
J (z) = J^a (z) \, t_a \, , \qquad (J^a)^* = \sum_b \tau_{ab} J^b
\ee
We want to investigate how these possible reality conditions for the
fields and the currents generalize to 
the case of a non-Abelian Toda 
model with $\wgs$ symmetry.

The action (\ref{eq1}) 
for the non-Abelian Toda model is defined for
an embedding $\cs \subset \cg$ 
which specifies the real Lie algebra $\cg_0$.
Now the the first term in the action 
$S_{\WZW}(g_0)$ makes sense whatever real form of 
$\cg_0^{\bf c}$ we choose, but it is natural to ask which of these
real forms is consistent with reality of the potential,
\be
\label{Todapot}
V = \tr \{ M_+ g_0 M_- g_0^{-1} \} 
\ee
and which of them arise from a choice of real form for $\cg^{\bf c}$.
The answer is provided by the following:

\begin{theo}
Consider a complex embedding 
$\cs^{\bf c} \subset \cg^{\bf c}$ with zero-grade 
subalgebra $\cg_0^{\bf c}$.
A real form $\cg_0$ for this subalgebra 
can be extended uniquely to a real form $\cg$ for the whole algebra 
which is compatible with the real embedding $\cs \subset \cg$ 
if and only if the Toda potential 
$\tr \{ M_+ g_0 M_-g_0^{-1} \}$ is real for any 
$g_0 \in G_0$.
\end{theo} 
Proof:
The implication is immediate in one direction---since for a real form
$\cg$ with subalgebras $\cs$ and $\cg_0$ 
the quantity (\ref{Todapot}) is manifestly real.

To show the converse is more involved. It is useful to begin
by thinking of $\cg_0$ as defined by a conjugation map $\tau_0$ on
$\cg_0^{\bf c}$. We want to show that we can extend this uniquely to a 
conjugation map $\tau$ on $\cg^{\bf c}$ which satisfies (\ref{sinv}).
We note first that if such a map exists it is certainly unique,
because any $x \in \cg^{\bf c}$ can be written in the form
$$
x = (\ad_{M_\pm})^m x_0 \ , \ \ x_0 \in \cg_0^{\bf c}
\qquad \Rightarrow \qquad 
\tau(x) = (\ad_{M_\pm})^m \tau_0 (x_0)
$$
in order to satisfy (\ref{sinv}).
We need to show that this actually defines 
a conjugation map on all of $\cg^{\bf c}$. 

Rather than doing this directly, however, the argument can be simplified
by focusing on elements at grades 0 and $\pm 1$.
We can make use of a result of Dynkin mentioned earlier, 
which states that for any
integral complex embedding $\cs^{\bf c} \subset \cg^{\bf c}$ we can choose a
Cartan subalgebra of grade zero elements and a basis of simple root 
generators with grades 0 or $\pm 1$. Since any
element of the algebra can be expressed in terms of these in a
canonical way, it is
enough to check that we have an automorphism acting on elements with
these specific grades. Thus it suffices to show that:
$$
[\tau(x),\tau(y)] = \tau([x,y])\mb{for} \left \{ \begin{array}{rl}
{\rm (i)} & x,y \in \cg^{\bf c}_0 \\
{\rm (ii)} & x\in \cg^{\bf c}_0,\,y \in \cg^{\bf c}_{\pm 1} \\
{\rm (iii)} & x\in \cg^{\bf c}_{1},\,y \in \cg^{\bf c}_{-1} \end{array} \right.
$$ 
In fact, we shall recast this in yet another equivalent form. 
Recall that the invariant subset of a conjugation map is a real Lie algebra
precisely because the conjugation map is an 
automorphism, as discussed in the last section.
If we consider the sets $\cg_n$ consisting of the 
elements fixed by $\tau$ at each specific grade,
then it is clear that the conditions above are equivalent to 
showing that:  
$$
[x,y] \in \cg \mbox{ for } \left \{ \begin{array}{rl}
{\rm (i)} & x,y \in \cg_0 \\
{\rm (ii)} & x\in \cg_0,\,y \in \cg_{\pm 1} \\
{\rm (iii)} & x\in \cg_{1},\,y \in \cg_{-1} \end{array} \right.
$$ 

\begin{itemize}
\item[(i)] True by assumption. 
\item[(ii)] Assume that (ii) is not true, 
i.e. that we can find $t_1 \in \cg_1$ 
and $t_0 \in \cg_0$ such that $[t_1,t_0] \not \in \cg_1$.
This implies that $[t_1,t_0]$, and for generic real $x$ also 
$e^{x t_0} t_1 e^{-x t_0}$, is a 
{\em complex} linear combination of elements of \cg. 

It is clear that any element in $\cg_1$ can be written in 
the form $\ad(M_+)(t_0),\,t_0\in\cg_0$. This implies that 
$\ad(G_0)(M_+)$ is dense in $\cg_1$ 
and we can therefore assume that we can 
find $g_0 \in G_0$ such 
that $t_1 = g_0 M_+ (g_0)^{-1}$. Collecting these statements, we find that 
$e^{x t_0} g_0  M_+ (g_0)^{-1} e^{-x t_0}$ is a {\em complex} 
linear combination 
of elements of \cg; it is now easy to find an element $g\in G_0$ 
such that 
$V = \tr \{ g M_+ g^{-1} M_- \} \not \in \fl{R}$, 
which is inconsistent with the assumptions. 
\item[(iii)] The proof of (iii) is very similar to the proof of (ii)
above. 
Assume (iii) does not hold, i.e. that we have $t_1 \in \cg_1$ 
and $t_{-1} \in \cg_{-1}$, such that 
$[t_1,t_{-1}] \not \in \cg$. This implies that 
$[t_1,t_{-1}]$ is a complex linear combination of elements in $\cg_0$. 

We can then find an element $t_0 \in \cg_0$ such that 
$$
\tr \{ [t_1,t_{-1}]  t_0 \} = \tr \{ [t_0,t_{1}]  t_{-1} \} \not \in \fl{R}
$$
and therefore, for generic real $x$, also   
$$
\tr \{ e^{x t_0}t_{1} e^{-x t_0} t_{-1} \} \not \in \fl{R}
$$
Using the same arguments as in (ii), we can assume that we can find 
$g_0$ and $h_0$ in $G_0$ such that 
$$
t_1 = g_0 M_+ (g_0)^{-1} \quad\quad t_{-1} = h_0 M_- (h_0)^{-1}, 
$$
and it is now easy to see that we can find a group element $g\in G_0$ such 
that 
$$
V = \tr \{ g M_+ g^{-1} M_- \} \not \in \fl{R}
$$
which contradicts the assumptions. 
\end{itemize}

We have shown that (i), (ii), and (iii) are true, and we conclude that 
$\tau$ is an automorphism; we have therefore proven theorem 2. 
\hfill {$\Box$} \\[3pt]

The result we have just proved 
generalizes an observation first made some time ago \cite{JE}
regarding the possible reality conditions for abelian Toda theories.

\begin{cor}
Consider a principal complex embedding $\cs^{\bf c} \subset \cg^{\bf
c}$ so that $\cg_0^{\bf c}$ is a Cartan subalgebra.
A conjugation map defining a compatible real form for 
$\cg^{\bf c}$ is given by a diagram automorphism, and the 
reality conditions for the Toda fields which ensure a real
lagrangian are therefore in one-to-one correspondence with symmetries 
of the Dynkin diagram.
\end{cor}

\subsection{Real Forms and W-algebras}
\label{realW}

The WZW model based on 
a real form $\cg$ of a complex Lie algebra $\cg^{\bf c}$
has a Kac-Moody symmetry 
$\widehat \cg$, which is in turn a real form of the complex Kac-Moody
algebra $\widehat \cg \otimes \fl{C}$.
In a similar way, 
an embedding 
$\cs \subset \cg$ gives rise to a real algebra
$\wgs$ which is a real form of the
complex extended conformal algebra
$\wgs \otimes \fl{C}$, which can be thought of as associated with 
the complex embedding $\cs^{\bf c} \subset \cg^{\bf c}$.
Although it is often convenient algebraically to work with this
complexified algebra, one should not
loose sight of the fact that it is some specific real form which
is truly the symmetry of the reduced WZW or non-abelian Toda theory.

For the Virasoro subalgebra of a $\cw$-algebra there 
is a standard set of hermiticity conditions which correspond to the
fact that the energy-momentum tensor of the theory is real in the
sense of (\ref{conj}).
In principle one can imagine the possibility of rather complicated real forms 
of a given complex $\cw$-algebra which might change these hermiticity
properties. In this paper, however, we shall always understand 
a \cw-algebra to come with a preferred Virasoro subalgebra, and we 
shall assume as part of its definition that a real form 
of a complex $\cw$-algebra always 
restricts to the standard real form for this preferred Virasoro
subalgebra. 
Since the generators of the Virasoro algebra $\{ L_{-1}, L_0, L_1 \}$ 
correspond to the preferred generators $\{M_-, M_0 , M_+ \}$ 
discussed earlier, any such real form corresponds to a compatible
conjugation map.
This really amounts to saying that  
for the complex algebras $\wgs \otimes \fl{C}$, the only real forms
we allow are of type $\wgs$.
 
Before we can state the main result of this section, we must set up
some notation.
In section 2.3 we described how the generators of $\wgs$ can be
associated with the irreducible representations occurring in the
decomposition of the adjoint for the embedding $\cs \subset \cg$.
If we introduce the same basis of generators $\{ t^\A_{(j,m)} \}$
for $\cg^{\bf c}$, where $\A$ labels irreducible representations of 
$\cs^{\bf c}$ as before, then the compatible conjugation map
$\tau$ which defines $\cg$ will in general 
act non-trivially. 
However, by virtue of (\ref{sinv}), the action must be of the form
\be\label{blocktau} 
\tau (t^{\A}_{(j,m)}) = \sum_\B t^{\B}_{(j,m)} \tau_{\B \A}
\ee
where the $(j,m)$ labels are unaffected and in particular 
$\tau_{\A \B}$ can have
non-zero entries only for representations $\alpha$ and $\beta$ of the
same total spin. 

\begin{theo}
Consider a real embedding $\cs \subset \cg$ defined from 
$\cs^{\bf c} \subset \cg^{\bf c}$ by a compatible conjugation map 
$\tau$. The generators of $\wgs$, labelled by $\alpha$, obey 
\be\label{wreal} 
(W^{\A})^* = \sum_{\B} \tau_{\A \B} W^{\B}
\ee
with $\tau_{\A \B}$ defined above in $(\ref{blocktau})$;
and in particular the energy-momentum tensor is always real.
Furthermore, the conformal weights of the generators of $\wgs$ are 
independent of the choice of real form $\cg$.
\end{theo}

\noindent
Proof:
Given the basis of Lie algebra generators in (\ref{blocktau}), 
the $\cw$-algebra 
generators $W^\A$ can be identified with the Kac-Moody currents
$J^{\A}_{(j,j)}$ in the highest-weight
DS gauge. The required condition (\ref{wreal}) follows from
(\ref{kmreal}) given (\ref{blocktau}).
Since the embedding is compatible with the real form, and the
energy-momentum tensor is associated with the generator $M_+$, it
follows that it is always real. 
Finally, the matrix for the conjugation map
in (\ref{blocktau}) only mixes $sl(2)$ representations of the same
spin. As explained in section 2, the spin determines the
conformal weight of the corresponding $\cw$-generator uniquely.
\hfill $\Box$\\[3pt]

We have mentioned that the algebra $\cg$ can always be recovered from 
$\wgs$ as its VPA. But we also 
know that there can be inequivalent embeddings in a
given $\cg$ which produce distinct \cw-algebras.
A manifestation of this is that when we consider a complex embedding
$\cs^{\bf c}\subset\cg^{\bf c}$ and then take real forms by using
compatible conjugation maps $\tau$ and $\tau'$, 
we may well find real embeddings 
$\cs \subset \cg$ and $\cs \subset \cg'$ which are inequivalent,
even when $\cg \cong \cg'$. The reason is that this isomorphism
of real algebras need not fix the embedded subalgebra $\cs$.
If the two real forms $\cg$ and $\cg'$ 
have zero-grade subalgebras $\cg_0$ and $\cg'_0$ and 
centralizers $\ck$ and $\ck'$, then a necessary condition for 
the isomorphism
$\cg \rightarrow \cg'$ to fix $\cs$ is that it should restrict to
isomorphisms 
$\cg_0 \rightarrow \cg'_0$ and $\ck \rightarrow \ck'$.
In fact it is quite easy to find examples where 
$\cg\cong\cg'$ and $\cg_0\cong\cg'_0$ but $\ck\not\cong\ck'$
(see section 5).
In this case the algebras $\wgs$ and $\cw^{ {\cal G}'}_{ {\cal S}}$
will certainly not be isomorphic. Moreover we see that 
there are inequivalent non-abelian Toda models
based on the {\em same\/} real Lie algebra $\cg_0$, but with 
Kac-Moody symmetries corresponding to different real forms of the same
complex algebra $\ck^c$.

One may wonder whether the real forms 
$\cg_0$ and $\ck$ are together enough to determine 
the non-abelian Toda model and its $\cw$-algebra completely. 
If $\cg_0$ and $\cg'_0$ are isomorphic 
they must have the same character, and 
we know from Corollary \ref{char3} (page \pageref{char3}), 
that the characters of $\cg$ and $\cg'$ are then also equal.
This means in most cases that 
$\cg$ and $\cg'$ are actually isomorphic (with the possible exceptions
of the few real forms mentioned in section 3.1 with 
degenerate values of the character ).
But we would like to know, in addition, whether this isomorphism
can always be chosen to fix $\cs$, in which case 
it would certainly be necessary to have $\ck$ and $\ck'$ isomorphic too. 
We shall not attempt to answer this question in all generality, but we
mention that for the wide variety of cases 
to be considered in the classification of section 5, 
the real forms $\cg_0$ and $\ck$ actually do specify the embedding 
$\cs \subset \cg$ and hence $\wgs$ uniquely.

A related point arises when we recall that most
embeddings $\cs^{\bf c}\subset\cg^{\bf c}$ are (with a numerable set of
exceptions) of the type $(\cg^{\bf c}, \ch^{\bf c})$. 
An important fact is that such an $sl(2,\fl{C})$ embedding is unique 
up to conjugation once the regular subalgebra 
$\ch^{\bf c}$ is given up to isomorphism.
The analogous result does not hold
for real embeddings, however.
If we consider real embeddings $(\cg,\ch)$ and $(\cg',\ch')$ 
defined by compatible conjugation maps 
$\tau$ and $\tau'$ as before, then 
our previous remarks imply that we may have 
$\cg\cong\cg'$ and $\ch\cong\ch'$, but $\ck\not\cong\ck'$.
Hence, specifying $\ch$ up to isomorphism does not determine (up to
conjugation) a generalized principal $sl(2,\fl{R})$ embedding.

Finally, it is also interesting to note that while a real form of an affine, or
infinite-dimensional, \cw-algebra always corresponds to a real form of
the corresponding finite \cw-algebra, the opposite statement is
false. 
In fact, let $T$ be the energy-momentum tensor in a given affine
algebra \cw~and denote by $\rho$ the map from \cw~to 
the corresponding finite algebra $\cw_{fin}$; 
then $\rho(T)$ is in the center of $\cw_{fin}$. 
This implies that we can have an automorphism
on a finite \cw-algebra which does not leave $\rho(T)$ invariant, and
therefore we may have a conjugation map of $\cw_{fin}$ which does not
descend from a conjugation map of $\cw$.  

\section{Real forms and non-abelian Toda theories \\ for 
algebras of rank 2} 
\label{sec-rank2}

We now illustrate some of our general results and show how
they can be used to find all integral embeddings into real 
Lie algebras of rank 2. 
These embeddings are each of the type $(\cg,\ch)$ which we
shall obtain by conjugation maps compatible with the 
complex embeddings $(\cg^{\bf c},\ch^{\bf c})$.
The regular subalgebras of any $\cg^{\bf c}$ are easily
found by first drawing its extended Dynkin diagram and then removing nodes to
obtain all the possible Dynkin diagrams for $\ch^{\bf c}$ \cite{FRS}. 
For the algebras $A_2$, $B_2 \cong C_2$ or $G_2$, the results
obtained by this procedure are shown in the table below.
Each black node denotes a short root, and in the last column we have
marked with a $*$ those embeddings $(\cg^{\bf c} , \ch^{\bf c})$ which
are integral. 

We shall examine each of the integral embeddings in detail below.
Although we could discuss the half-integral embeddings in a similar
way, these are of less direct interest from the point of view of
integrable Toda theories (at least of the simplest kind).
The second and third embeddings for both $B_2$
and $G_2$ are not really distinct. It is a well-known feature of Dynkin's
method that it can give rise to embeddings which are conjugate to one
another, even when the regular subalgebras concerned are clearly not
conjugate, and this is what happens in these two examples. 
Representations of Lie algebras will be denoted, as usual, by their
dimension, except that for $A_1$ we shall also write $(j)$ to mean
the irreducible representation of spin $j$.

\begin{tabular}{|l|l|l|l|l|l|}
\hline
$\cg^{\bf c}$ & Dynkin Diagram & Extended Diagram & Sub-Diagrams &
$\ch^{\bf c}$ & \\ 
\hline \hline
$A_2$ & \Atwo & \Atwoe & \Atwo & $A_2$ &$*$ \\ 
     &       &        & \Aone  & $A_1$ & \\ \hline  
$B_2$ & \Ctwo & \Ctwoe & \Ctwo & $B_2$ &$*$  \\
      &       &        & \twoAone & $A_1 \oplus A_1$ &$*$   \\         
      &       &        & \Aonex & $A_1$ &$*$    \\ 
      &       &        & \Aone  & $A_1$ &   \\ \hline          
$G_2$ & \Gtwo & \Gtwoe & \Gtwo  & $G_2$ &$*$    \\
      &       &        & \Atwo  & $A_2$ &$*$    \\         
      &       &        & \twoAonex & $A_1 \oplus A_1$ &$*$   \\         
      &       &        & \Aone  & $A_1$  &\\         
      &       &        & \Aonex & $A_1$  &\\ \hline 
\end{tabular}\\[5mm]  

\subsection{Embeddings in $A_2$}

The only integral embedding in $A_2$ is the principal one 
\be\label{aprinc}
(A_2,A_2): \qquad
M_0 = h_1 + h_ 2 , \qquad M_\pm = 
\sqrt{2} (e_{\pm \alpha_1} + e_{\pm \alpha_2})
\ee
The decomposition of the adjoint representation is 
$
{\bf 8} \rightarrow (2) \oplus (1) 
$
and the zero-grade subalgebra is simply the Cartan subalgebra $U_1 \oplus U_1$.
This reduction corresponds to the Abelian Toda theory of type $A_2$.
Introducing fields 
$$ \Phi = \phi_1 h_1 + \phi_2 h_2 
$$ to parametrize the zero-grade subalgebra, we 
see that the WZW part of the action
is essentially trivial, while the potential term reads
$$
V = e^{-2\phi_1+\phi_2}+e^{-2\phi_2+\phi_1} 
$$
The $\cw$-algebra of the theory is generated by the 
energy-momentum theory $T$ of spin 2 and one additional primary $W$ of
spin 3, these being even and odd respectively under the interchange
$\phi_1 \leftrightarrow \phi_2$.
This is the famous $\cw_3$ algebra of Zamalodchikov.

The split real form of $A_2$ is
$sl(3,\fl{R})$ corresponding to the identity in the Weyl group and 
a conjugation map
which fixes all the Cartan-Weyl generators. The fields $\phi_j$ and
the generators $T$ and $W$ are 
then all real.
There is only one other conjugation map (up to isomorphism)
which is compatible with the embedding (\ref{aprinc}).
It is given by the root automorphism
$\tau: \A_1 \leftrightarrow \A_2$ which acts as a reflection symmetry
of the Dynkin diagram. This defines the real form $su(2,1)$.

When we impose the conditions 
(\ref{real}) corresponding to the new real form we find 
\be
\label{realsl3}
(\phi_1)^* = \phi_2 \qquad \iff \qquad 
\phi_1 = u + i v \, ,\quad \quad \phi_2 = u - i v 
\ee
with $u$ and $v$ real. 
The Toda potential is 
$$
V = 2e^{- u}\cos(3 v)
$$
when written in terms of the new fields, which is manifestly real, in
keeping with Theorem 2.
By virtue of their behaviour under $\phi_1 \leftrightarrow \phi_2$,
the generators of the $\cw$-algebra now obey
$$
T^* = T \quad \quad W^* = -W
$$
in accordance with Theorem 3.
This constitutes a new real form of the algebra $\cw_3$ which has in fact 
been considered previously in \cite{Hon}. 

\subsection{Embeddings in $B_2$} 

The positive roots of $B_2$ are $\A_1$ (long), $\A_2$ (short), 
$\A_1+\A_2$ and $\A_1+2\A_2 = \psi$, the highest root.

The principal embedding is integral, as always, and results in an abelian
Toda model. The adjoint representation decomposes as
${\bf 10} \rightarrow (3) + (1) $
and the corresponding $\cw$-algebra therefore has a single generator 
of spin 4 in addition to the energy-momentum tensor.
The standard reality conditions for this Toda theory correspond to the 
maximally non-compact form $so(3,2)$.
There is no other real form of $B_2$ compatible with the principal
embedding (by corollary 2, since there is no symmetry of the Dynkin
diagram) and we omit further discussion of this case. 

Now we turn to the more interesting case of non-principal embeddings.
Consider first the embedding defined by the short root
$$
(B_2, A_1): \qquad M_0 = h_2 , \qquad M_\pm = \sqrt{2} e_{\pm \alpha_2}
$$
The decomposition of the adjoint is
$$
{\bf 10} \rightarrow (1) \oplus (1) \oplus (1) \oplus (0)
$$
and the zero-grade subalgebra is 
\be\label{bzg}
A_1 \oplus U_1 = \{ h_1{+}h_2, \sqrt{2} e_{(\alpha_1 + \alpha_2)}, 
\sqrt{2}e_{-(\alpha_1 + \alpha_2)} \} \oplus
\{h_2 \}
\ee
The other integral embedding in the table is defined 
by a pair of orthogonal long roots
$$
(B_2, A_1 \oplus A_1): \qquad M_0 = h_2, \qquad M_\pm = 
e_{\mp \alpha_1} + e_{\pm \psi}
$$
In this case the adjoint representation decomposes via representations of 
$A_1 \oplus A_1$ in the pattern
$$
{\bf 10} \rightarrow (1,0) \oplus (0,1) \oplus (1/2,1/2) \rightarrow 
(1) \oplus (1) \oplus (1) \oplus (0)
$$
The zero-grade subalgebra is obviously again given by (\ref{bzg}).
The coincidence of the final pattern of representations and the 
zero-grade subalgebras confirm our assertion that these embeddings
are equivalent; in fact it can be shown that they are conjugate 
(by exponentials involving the generators 
$e_{\pm (\alpha_1 + \alpha_2)}$) but
we shall omit the details.
The pattern of representations reveals that 
we have a $\cw$-algebra consisting of
an abelian Kac-Moody generator $U$ of spin 1 and two primary fields 
$W^\pm$ of spin 2 in addition to the energy-momentum tensor $T$.

An efficient and quite general way to calculate the Toda potential is to 
first identify the representations of the zero-grade subalgebra to which the
generators $M_\pm$ belong, since the potential term involves precisely
the adjoint action of this subalgebra on these generators.
In the present case it is clear that the sets $\{ e_{\mp \alpha_1} , e_{\pm
\alpha_2} ,  e_{\pm(\alpha_1 + 2 \alpha_2)} \}$ carry spin-1
representations of the $A_1$ factor in (\ref{bzg}) 
and that they have opposite weight under 
the $U_1$ generator.
The potential term can now be readily calculated for either of 
the descriptions $(B_2,A_1)$ or $(B_2,A_1 \oplus A_1)$
given above (some details are given in appendix B). 
Introducing a parametrization
\be\label{bfields}
\Phi = u M_0 + 2\phi_0 (h_1{ + }h_2) 
+ \sqrt{2}\phi_+ e_{\alpha_1 + \alpha_2} 
+ \sqrt{2}\phi_- e_{- (\alpha_1 + \alpha_2)} 
\ee
for $\cg_0$ and taking $g = \exp \Phi$ we find (up to an overall constant)
\beano
(B_2, A_1): && V = 
\frac{e^{- u}}{\rho^2} ( \phi_0^2  + \phi_+ \phi_- \cosh 2\rho ) \nn
(B_2, A_1 \oplus A_1): && V =
\frac{e^{- u}}{4 \rho^2}  
\left \{ \left (  (\phi_+{-}\phi_-)^2 - 4\phi_0^2 \right ) \cosh 2\rho
- (\phi_+{+}\phi_-)^2 \right \}  \nn
\enano
where $\rho^2 = \phi_0^2 + \phi_+\phi_-$. 
The apparent difference in these potentials is really just an
artifact of the way we are parametrizing $\cg_0$; they are related by 
exchanging $(\phi_+ + \phi_-) \leftrightarrow 2 \phi_0$. 
This corresponds precisely to the action on $\cg_0$ 
of the isomorphism on $\cg$ which relates the embeddings.

Let us now consider the possible real forms consistent with these
embeddings. For the split real form 
$\cg = so(3,2)$ the Toda fields and the
$\cw$-algebra generators discussed above are all real and 
the zero-grade subalgebra is $\cg_0 = su(1,1) \oplus gl(1)$.
There is only one other real form which is compatible.
This can be defined by the conjugation map 
$$
\tau (h_1) =  - (h_1 + 2h_2) , \qquad \tau (h_2) = h_2 , \qquad
\tau (e_{\pm \alpha_1}) = e_{\mp (\alpha_1 + 2 \alpha_2)} ,
\qquad \tau (e_{\pm \alpha_2}) = e_{\pm \alpha_2}
$$
which in the $(B_2 , A_1 \oplus A_1)$ description simply amounts to
exchanging the two $A_1$'s. 
The resulting real forms are $\cg = so(4,1)$
and $\cg_0 = su(2) \oplus gl(1)$, rendering the non-abelian factor
compact in the latter case.

In terms of the parametrization (\ref{bfields}) the effect,
according to eq. (\ref{real}), is to take 
$$(\phi_0)^*=-\phi_0 \quad
(\phi_\pm)^*=-\phi_\mp
\qquad 
\iff 
\qquad
\phi_0 = i \phi_3, \quad \phi_{\pm} = \pm \phi_1 + i \phi_2 
$$ 
with $\phi_1$, $\phi_2$ and $\phi_3$ real.
The Toda potentials can now of course be written in terms of 
these new real fields. For the $(B_2,A_1)$ case we find 
$$
V = \frac{e^{-u}}{\phi^2} (  
(\phi_1^2+\phi_2^2)\cos 2\phi + \phi_3^2 ) 
$$
with ${\phi}^2 = \phi_1^2+\phi_2^2+\phi_3^2$, while for the
$(B_2,A_1 \oplus A_1)$ embedding 
we have $\phi_2 \leftrightarrow \phi_3$ in the above, in keeping
with the relationship between the coordinate systems mentioned earlier. 

The reality properties of the \cw-algebra now change to
$$
T^* = T , \quad U^* = -U , \quad (W^\pm)^* = \pm W^\pm
\iff 
U = i \tilde{U}, \quad W^\pm = \pm W^1 + i W^2
$$ 
where $\tilde{U}$ and $W^1$,$W^2$ are real. The detailed form of
this classical $\cw$-algebra can be found in our previous paper \cite{EM}. 
The representation theory of the algebra has also been analyzed in
\cite{PB}. 
It is interesting that it is the second (non-split) 
real form of the group which results in a compact (though abelian)
Kac-Moody symmetry, and which is therefore the real form of the 
$\cw$-algebra possessing 
highest-weight representations of the standard kind. 

\subsection{Embeddings in $G_2$}

The positive roots of $G_2$ are $\A_1$ (long), $\A_2$ (short), 
$\A_1+\A_2,\,\A_1+2\A_2,\,
\A_1+3\A_2$ and $2\A_1+3\A_2 = \psi$, the highest root.

For the principal embedding
the adjoint decomposes as
${\bf 14} \rightarrow (5) \oplus (1)$
which gives rise to an abelian Toda model whose $\cw$-algebra contains
a primary field of spin 6 in addition to the energy-momentum tensor.
Once again there are no non-trivial choices for a conjugation map in 
this case, since there is no symmetry of the $G_2$ Dynkin diagram.

There are two superficially different 
non-principal integral embeddings listed in the table.
First we have 
$$
(G_2,A_2): \qquad M_0 = - (h_1 + 3h_2) ,
\qquad M_\pm = \sqrt{2} ( e_{\pm \A_1} + e_{\mp \psi} )
$$
for which the adjoint of $G_2$ decomposes via representations of $A_2$ as:
$$
{\bf 14} \rightarrow {\bf 8} \oplus {\bf 3} \oplus \bar {\bf 3}
\rightarrow (2) \oplus (1) \oplus (1) \oplus (1)
$$
Alternatively we have 
$$
(G_2,A_1 \oplus A_1) : \qquad M_0 = -(h_1 + 3h_2) ,
\qquad M_{\pm} =  e_{\pm \alpha_1} + \sqrt{3} e_{\mp (\alpha_1 + 2\alpha_2)} 
$$
for which the adjoint decomposes via representations of $A_1 \oplus
A_1$ as 
$$
{\bf 14} \rightarrow (3/2,1/2) \oplus (1,0) \oplus (0,1) 
\rightarrow (2) \oplus (1) \oplus (1) \oplus (1)
$$
In both cases the zero-grade subalgebra is 
\be\label{gzg}
A_1 \oplus U_1 = \{ \sthalf (h_1{+}h_2) , 
\sqrt{3}e_{\alpha_1 + \alpha_2 } , 
\sqrt{3}e_{-(\A_1+\A_2)} \} \oplus \{ h_1 + 3h_2 \}
\ee
As in the example of $B_2$, these embeddings can be shown to be
conjugate in $G_2$ but we shall not discuss the details. For
definiteness, we deal from now on with the $(G_2,A_2)$ description.

From the decomposition of the adjoint representation we see that the
resulting $\cw$-algebra has primary fields of spins 3, 2, 2 in
addition to the energy-momentum tensor.
As in the $B_2$ example, we can calculate 
the Toda potential by identifying the
representations to which $M_\pm$ belong.
It is easy to see that 
$\{ e_{\mp\alpha_1} , e_{ \pm\alpha_2} , e_{ \pm (\alpha_1 + 2\alpha_2)} ,
e_{ \pm (2 \alpha_1 + 3 \alpha_2)} \}$ form spin-3/2 representations.
Adopting a similar explicit parametrization for the zero-grade
subalgebra as in the $B_2$ case 
(\ref{bfields}) we find the Toda potential (up to a numerical constant)
\be
V = {e^{-u}\over \rho^3}\left \{ 2 \rho^3\cosh^3\rho + 
6 \phi_0^2 \rho \cosh\rho \sinh^2\rho  + 
(\phi_+^3+\phi_-^3) \sinh^3 \rho \right \}
\ee
with $\rho^2=\phi_0^2+\phi_+\phi_-$ (see appendix B for further details). 

Turning our attention now to the possible real forms, we have as usual
the maximally non-compact form for $G_2$, in
which case the Toda fields and the $\cw$-algebra generators referred
to above are all real. To look for other possibilities compatible with
the non-principal embedding, we seek a root automorphism
$\tau \in \mbox{Aut}(\Delta)$ which could be used 
to construct a conjugation map 
$\tau(e_\A)=\chi_\A e_{\tau(\A)}$ and $\tau(h_\A)=h_{\tau(\A)}$, 
with $\chi_\A=\pm 1$, as explained in the previous section.
To leave $M_+$ invariant we have just two possibilities.
Either $\tau$ fixes both $\A_1$ and $\psi$, in which case we quickly 
reduce to the trivial case, 
or else it exchanges them:
$\tau(\A_1) = -\psi$ and $\tau(\psi) = - \A_1$ (this is 
reflection in the root $\A_1+\A_2$). 
The constants $\chi_\A$ must satisfy the consistency conditions 
(\ref{def-chi}), which leads to 
$\chi_{-\psi} = -\chi_{\A_2}$ and 
$\chi_{\A_1+\A_2} = -\chi_{\A_1}\chi_{\A_2}$. 
We must ensure $\tau (M_+) = M_+$ which implies that 
$\chi_{\A_1} = \chi_{-\psi} = 1$ and therefore $\chi_{\A_1+\A_2} = 1$. 
But it then turns out (for example by computing characters) that 
this conjugation map again defines the split real form of $G_2$.
Despite appearances, therefore, there is actually no other real
form compatible with the non-principal embedding, and hence no
alternative real form of the $\cw$-algebra. 

It is instructive to reconcile these observations with our
formula for the potential.
The reality conditions corresponding to the 
automorphism $\tau$ which exchanges $\A_1$ and $\psi$ 
would be $(\phi_0)^* = -\phi_0$ and 
$(\phi_1)^* = \sigma \phi_2$, where an easy calculation shows that in
fact $\sigma = \chi_{\A_1+\A_2}$. Taking $\sigma=+1$ 
leads once again to the split real form. If we try to take $\sigma=-1$, 
however, we see that this choice leads to a potential with a non-zero
imaginary part. The fact that this real form is not compatible with
the embedding is therefore entirely consistent with our earlier
Theorem 2.

\section{Embeddings (\cg,\ch) with \cg~classical and \ch~simple} 
\label{sec-3} 

In this section we describe the classification of real forms
compatible with integral $sl(2,\fl{C})$ embeddings of the type 
$(\cg^{\bf c},\ch^{\bf c})$ where
$\cg^{\bf c}$ is one of the classical 
algebras and $\ch^{\bf c}$ is a simple,
regular subalgebra.
The general strategy which we outline below 
could in principle be applied to any integral
embedding in a regular subalgebra. 
Embeddings in the exceptional algebras could
certainly be dealt with in a similar fashion and 
there is also nothing technically which constrains us to consider only
simple subalgebras, although this is
clearly the most natural case to investigate first 
and there would be a vast proliferation of
additional examples for non-simple subalgebras.

The first step is to find concrete expressions for 
root automorphisms and conjugation maps 
which we can use to define all the real forms 
of the classical algebras. Once we have done this, the next step will
be to examine each of the simple regular subalgebras 
which correspond to integral
embeddings and then determine the subset of conjugation maps 
which leave invariant the relevant principal embedding. 
This second step is greatly simplified by the fact 
that the choice of subalgebra $\ch^{\bf c} \subset \cg^{\bf c}$ 
fixes the complex embedding uniquely up to conjugation.
This means that we can make convenient assumptions about which simple
roots of $\cg^{\bf c}$ are also simple roots of $\ch^{\bf c}$ with no
loss of generality. We shall also find it useful to 
describe each embedding by giving its equivalent {\em characteristic diagram},
for which one assigns the grade 0 or $\pm 1$ to each simple root 
of $\cg^{\bf c}$.
The third and last step is to calculate the resulting real forms $\cg_0$ and 
$\ck$ which are the aspects of the embedding most directly 
associated with the properties of the non-abelian Toda theory. 

In the following subsections we shall discuss in turn 
embeddings in $A_n$, $D_n$, and then the related cases $B_n$ and $C_n$.
We shall certainly not include every detail of the arguments, 
since this would make our account prohibitively lengthy. 
We shall also take the liberty of
omitting more and more routine technicalities as we work our way
through the list of possibilities.
The final results of the classification are collected together as Theorem 5
and summarized in the table on page \pageref{newtable1}.  

\subsection{Conjugation Maps and Root Spaces} 

Recall that an automorphism of $\ch^{\bf c}$ which fixes a principal
$sl(2,\fl{C})$ subalgebra must fix the Cartan subalgebra 
$H_{\cal H}$ and must furthermore be a diagram automorphism of  
$\ch^{\bf c}$.
We therefore seek all automorphisms of $\cg^{\bf c}$ which 
act as diagram automorphisms on some regular
subalgebra $\ch^{\bf c}$.
We claim that it is sufficient to consider automorphisms 
that fix a Cartan subalgebra $H_{\cal G}$ of the larger algebra
$\cg^{\bf c}$. 
This can be regarded as a slight generalization 
of the well-known result that any
automorphism of a complex Lie algebra 
is conjugate by an inner automorphism 
to one which fixes a given Cartan
subalgebra. 

\begin{theo}
Consider a complex simple Lie algebra $\cg^{\bf c}$ 
and a regular subalgebra $\ch^{\bf c}$.
Choose Cartan subalgebras $H_{\cal G}$ and $H_{\cal H}$ respectively with
$H_{\cal H} \subset H_{\cal G}$.
Any automorphism of $\cg^{\bf c}$ which restricts to an
automorphism of $\ch^{\bf c}$ and which fixes
$H_{\cal H}$ is conjugate to an automorphism 
with all these properties but which in addition fixes $H_{\cal G}$.
\end{theo}

\noindent
Proof: Define $\cz^{\bf c}$ to be the centralizer of $\ch^{\bf c}$ 
in $\cg^{\bf c}$. We may choose a Cartan subalgebra
$H_{\cal Z}$ for this centralizer with the property that 
$H_{\cal G} \subset H_{\cal H } + H_{\cal Z}$
(in fact the spaces 
$H_{\cal H}$ and $H_{\cal Z}$ must be orthogonal).
Let $\tau$ be the automorphism fixing $H_{\cal H}$ and $\ch^{\bf c}$. 
We will show that $\tau$ is conjugate to an automorphism $\tilde \tau$ which
fixes not only $H_{\cal H}$ but also $H_{\cal Z}$, 
which is clearly enough to show that it fixes $H_{\cal G}$. 
First note that $\tau$ fixes 
$\cz^{\bf c}$, simply because it also fixes 
$\ch^{\bf c}$.
But this means that $\tau$ restricts to an automorphism on
$\cz^{\bf c}$ and hence that there is an inner automorphism $\sigma$
on $\cz^{\bf c}$ such that $\sigma^{-1} \tau \sigma$ fixes 
$H_{\cal Z}$. Since inner-automorphisms are nothing but conjugation
by some group element,
$\sigma$ is actually automatically defined as an inner-automorphism on all
of $\cg^{\bf c}$.
Furthermore, $\sigma$ restricts to the identity on 
$\ch^{\bf c}$ precisely because $\cz^{\bf c}$ is its 
centralizer. 
The desired automorphism of $\cg^{\bf c}$ is therefore 
$\tilde \tau = \sigma^{-1} \tau \sigma$.
\hfill $\Box$\\[3mm]
 
To write down conjugation maps we follow the procedure
explained in section 3 for constructing algebra automorphisms from
the root automorphisms Aut($\DE$).
The root automorphisms are in turn conveniently described
by regarding the roots as living in $\fl{R}^m$ 
with its standard basis vectors 
$\{e_i \}$ obeying 
$e_i \cdot e_j = \D_{ij}$ where $i, j = 1 , \ldots , m$.
For the algebras $B_n,\,C_n$ and $D_n$ we take $m=n$ so that this is
exactly the span of the set of roots $\DE$. 
For $A_n$, however, we take $m=n+1$, and the span of $\DE$ is the 
subspace of co-dimension one orthogonal to $\sum_{i=1}^{n+1} e_i$.
The precise definitions of the roots are given below.

The important point for our purposes 
is that the group Aut($\DE$) always acts as a 
subgroup of the {\em signed permutations\/} of the basis vectors 
$\{ e_i \}$; 
by which we mean permutations accompanied by a change of sign of any
number of the basis vectors.
We can of course decompose any permutation into disjoint cycles, and 
we shall use the phrase $m$-cycle to mean one of length or order $m$.
We shall also refer to an {\em inverted\/} $m$-cycle, meaning an
$m$-cycle accompanied by a change of sign of all the basis vectors it
permutes. Thus the map
$\tau(e_i) = e_{i+1}$ for $i=1,2,\ldots,m{-}1$ and 
$\tau(e_m) = e_1$ is a typical $m$-cycle, while 
$\tau(e_i) = -e_{i+1}$ for $i=1,2,\ldots,m{-}1$ and  
$\tau(e_m) = -e_1$ is a typical inverted $m$-cycle. 
Note that a 1-cycle is trivial as a permutation, but an inverted
1-cycle is non-trivial because it reverses the sign of a single basis vector.
The elements of Aut($\DE$) of order two are precisely the
permutations that consist entirely of 1-cycles and 2-cycles
(transpositions) each with or without inversions. 

\subsection{Embeddings in $A_n$}
\label{An}

The root space of $A_n$ is the subspace of 
$\fl{R}^{n+1}$ orthogonal to $\sum_{i=1}^{n+1} e_i$.
The set of roots is 
$\DE = \{ e_k - e_l \, : \, k\not=l \}$ and the simple roots are  
$\A_k=e_{k} - e_{k+1}$ with $k=1,\ldots,n$. 
The Weyl group $W(\DE)$ is the group of permutations of the 
vectors $e_i$; Aut($\DE$) is the group of permutations with a possible
sign-change of {\em all} basis vectors. Note that the quotient is then
indeed $\fl{Z}_2$, the symmetry of the Dynkin diagram.

\subsubsection{Real forms of $A_n$} 
\label{rfAn} 

We will show how to construct all possible real forms of
$A_n$ by defining conjugation maps from the 
elements of Aut($\DE$) as in eq.~(\ref{Waut}). 

Consider a conjugation map constructed from an element $\tau$ of 
$W(\DE)$. Suppose that $\tau$ can be decomposed into $p$ 2-cycles
and $n{+}1{-}2p$ invariant basis vectors. 
One such map is:
\be
\label{tauA}
\begin{array}{rcll}
\tau(\A_{2i-1}) & = & -\A_{2i-1}  &  1 \le i \le p \\
\tau(\A_{2i}) & = & (\A_{i-1}+\A_i+\A_{i+1})  &  1 \le i < p \\
\tau(\A_{2p}) & = & (\A_{2p-1}+\A_{2p}) \\
\tau(\A_{i}) & = & \A_{i} & i > 2p
\end{array}
\ee
We want to calculate the characters of the possible real forms which
result from this by using equation (\ref{char2}) of Theorem 1.
The trace of
$\tau$ restricted to the Cartan subalgebra is $\mbox{tr}(\tau_H) = n{-}2p$.
Next we consider those 
roots which satisfy $\tau(\A) = - \A$, namely 
$\A_1,\,\A_3,\ldots,\A_{2p-1}$. 
The parameters $\chi_\alpha$ which enter in equation (\ref{char2})
are strongly restricted in this case by the consistency 
condition $\chi_\A = \chi_{\tau(\A)}$, which implies 
$\chi_{\A_1} = \chi_{\A_3} = \cdots = \chi_{\A_{2p-1}}\equiv\chi$. If
$2p = n+1$ then we may choose $\chi=\pm 1$, but if $2p \not = n+1$ then
we have the unique possibility $\chi=1$. 
The result of the calculation for the character is therefore 
$$
\sigma = n-2p + 2 \chi p =   
\left \{ \begin{array}{rl} -n-2 & \mbox{ if } 2p = n+1 \mbox{ and } 
\chi = -1 \\ 
n & \mbox{ all other cases} \end{array} \right.
$$

From the table of real forms and characters (\ref{tab0})
we see that any element of the Weyl group 
which leaves a basis vector invariant can
give rise only to conjugation maps defining 
the split real form $sl(n{+}1)$ of $A_n$. The only elements of the
Weyl group capable of producing a different real form are those consisting
entirely of 2-cycles. This is only possible when $n$ is odd, and the result
is the real form $su^*(n{+}1)$. To obtain the remaining real forms of $A_n$ we
must use root automorphisms which are not elements of the Weyl group.

Consider an element $\tau\in$ $\mbox{Aut}(\DE)\setminus W(\DE)$ which
consists of $p$ inverted 2-cycles and $n{+}1{-}2p$ inverted basis vectors. 
One such $\tau$ has an action on the simple roots just as in 
(\ref{tauA}) except for a change of sign on the right hand
side of every equation. Now we have $\mbox{tr}(\tau_H) = -(n{-}2p)$.
The roots that satisfy $\tau(\A)=-\A$ are now the roots in the 
$A_{n-2p}$ subalgebra generated by the $n+1-2p$ inverted
basis vectors. For each positive root of this subalgebra 
$\alpha \in \DE_+(A_{n-2p})$ we can consistently choose $\chi_\A=-1$ 
and then we find the character 
$$
\sigma = 2p-n - 2 |\DE_+(A_{n-2p})| = - \mbox{dim}(A_{n-2p})
= 1 - (n+1 -2p)^2
$$
Comparing this result with table \ref{tab0} we find that these
are exactly the characters of $su(n\!+\!1\!-\!p,p)$ for
$p=0,\ldots,[\frac{n+1}{2}]$. 

\subsubsection{Invariant regular subalgebras}
The simple, regular subalgebras of $A_n$ are $A_m$ with $m\le n$. 
The embedding is integral when $m$ is even or when $m = n$ is odd,
the latter giving the principal embedding.
The embedding is half-integral---and hence irrelevant for our 
purposes---for odd $m < n$.

We may describe a grading by assigning a number $j_k$ to 
each basis vector $e_k$, so that the grade of any root 
$e_k-e_l$ is $j_k-j_l$. 
It is then simple to identify the zero-grade subalgebra
as the set of all Cartan generators together with those
step-operators for roots $e_k-e_l$ with $j_k = j_l$. 
We can alternatively define a grading by specifying an integer
$p_i$ for each simple root $\A_i$ and we denote this on the Dynkin
diagram by writing $p_i$ above the appropriate node.
For example, we can take an embedded $A_m$ subalgebra 
generated by the first $m$ simple roots of $A_n$, and then the grading
corresponds to the diagram: 
$$
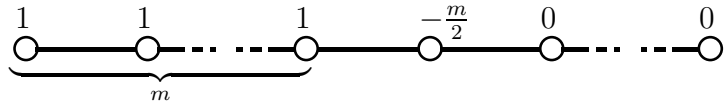
\underbrace{\cir{1}\li\cir{1}\hlih\cir{1}}_m\hspace{-.4mm}
\li\cir{$-\frac{m}{2}$}\li\cir{0}\hlih\cir{0}
$$
It is always possible to make a Weyl transformation to a 
basis where the gradings of the simple roots are 0 or $\pm 1$; 
the resulting characteristic 
Dynkin diagram \cite{Dy} is: 
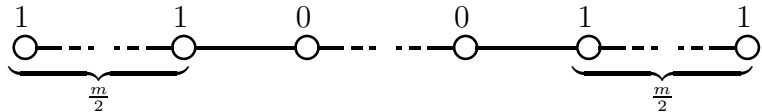
$$
\underbrace{\cir{1}\hlih\cir{1}}_{\frac{m}{2}}\hspace{-.5mm}\li
\cir{0}\hlih\cir{0}\li\hspace{-.5mm}
\underbrace{\cir{1}\hlih\cir{1}}_{\frac{m}{2}}
$$
Note that in the case of the principal embedding of $sl(2,\fl{C})$ 
into an algebra \cg, (\cg,\cg) in our notation, the grades of all
simple positive roots are 1 by definition. 

Now the zero-grade subalgebra is generated by all Cartan
generators and by the step operators corresponding to simple
roots that have grade zero.
In this case we see that $\cg^{\bf c}_0 = A_{n-m} \oplus m U_1$. 
It will be important that exactly one of the basis
elements $e_i$ is common to the root-spaces of {\em
both} $A_m$ and $\cg^{\bf c}_0$; this is 
$e_{{m\over 2}+1}$ in the basis where
$A_m$ is generated by the first $m$ simple roots of $A_n$. 

We wish to consider automorphisms of $A_n$ which leave $A_m$ invariant
up to a diagram automorphism. 
In the previous subsection we found that elements of $W(\DE)$ 
lead only to the real forms 
$sl(n\!+\!1)$ or $su^*(n+1)$.
This first real form can of course be realized by the identity
in the Weyl group, which certainly leaves the $A_m$ subalgebra 
invariant. For the second real form, however, 
we saw above that the corresponding elements of the Weyl group 
must consist entirely of 2-cycles, and this implies that
the only regular subalgebra it can leave invariant is one 
of type $A_1$.
In fact if we now consider more general root automorphisms in 
$\mbox{Aut}(\DE) \setminus W(\DE)$,
then we find likewise that 
the only simple, regular subalgebras that can be invariant 
are again those of type $A_1$.
We conclude that only the split real form of $A_n$ corresponds to a
conjugation map which restricts to the identity on $A_m$. This 
clearly results in the split real form of the zero-grade subalgebra.

The remaining possibility which we must investigate 
is that the conjugation map on
$A_n$ acts as a {\em non-trivial} diagram automorphism on $A_m$.
Any such non-trivial diagram automorphism $\tau$ 
acts on $A_m$ by interchanging 
$e_{i}$ and $-e_{m+2-i}$, i.e. it consists of 
$\frac{m}{2}$ inverted 2-cycles and
one inverted basis vector $e_{m+2 \over 2}$.
 
Let us now turn to the action of such an automorphism on 
$\cg^{\bf c}_0 = A_{n-m} \oplus m U_1$. 
The abelian part $m U_1$ is composed of the Cartan generators 
of $A_m$ and the action of $\tau$ on
it is therefore completely determined;  
the resulting real form is 
$\frac{m}{2} gl(1) \oplus \frac{m}{2} u(1)$.  
When we come to the non-abelian part, we find that we 
can choose freely the action of $\tau$ on the
basis vectors of the root-space of $A_{n-m}$, except for the fact that
it must invert the
basis vector $e_{{m\over 2}+1}$ which, as we recall, lies 
in the root-spaces of both $\cg^{\bf c}_0$ and $A_m$.  
We conclude that the only restriction on the real form
of the simple part of $\cg^{\bf c}_0$ is that it be constructed using an
automorphism which inverts at least one basis vector. In other words,
it cannot consist entirely of (inverted) $2$-cycles.  
From the results of the previous subsection, we find that by using such 
elements of Aut($\DE$) 
we can define conjugation maps giving each of the real forms 
$su((n{-}m){+}1{-}p,p)$ with $p=0,\ldots, [ \frac{n-m+1}{2} ]$ of 
$A_{n-m}$. 
It is not possible, however, to get the real form $su^*(n{-}m{+}1)$. 

Finally we come to the centralizer which is of complex type
$\ck^{\bf c}= A_{n-m-1}\oplus U_1$.
The split real form of $A_n$,
corresponding to the identity map on $A_m$, obviously 
results in the real form 
$\ck = sl(n{-}m) \oplus gl(1)$ for the centralizer.
Aside from this, however, we must consider the broader set of 
possibilities arising from non-trivial diagram automorphisms of $A_m$. 

We first note that the abelian factor $U_1$ in $\ck^{\bf c}$ is
generated by the projection of the vector $\sum_{i=1}^m e_i$. 
It follows that any non-trivial diagram automorphism on $A_m$ produces
the compact form $u(1)$ of this abelian factor.
Now the relevant real forms of the simple, non-abelian factor 
$A_{n-m-1}$ are clearly
$su((n{-}m){-}r,r)$ with $r=0,\ldots, [ \frac{n-m}{2} ]$.
It is quite possible for a given real form $\cg_0$ to give rise to 
a number of different real forms $\ck$. In other words, there may
be more than one value of $r$ in this list which is allowed for a
given value of $p$ in our earlier list of real forms $\cg_0$.
Nevertheless, the values of $r$ are severely restricted simply by the fact 
that $\ck \subset \cg_0$. In the present case it is clear that this is
satisfied only if $r= p$ or $r = p{-}1$ (assuming ranges of values
where these equations make sense).

To show that both these possibilities can indeed occur, we can exhibit
the relevant conjugation maps. 
Set $k=n{-}m$ for convenience and take the simple subalgebras
$A_{k-1} \subset \ck^{\bf c}$ and 
$A_k \subset \cg_0^{\bf c}$ 
to be generated by the first $k$ and first $k{+}1$ basis vectors 
respectively.
As earlier, we consider an
automorphism $\tau$ on $\cg_0^{\bf c}$ which inverts the first $k{-}2p$ 
basis vectors, and consists of inverted 2-cycles for the rest. We choose 
$\chi_{\A_2}=\chi_{\A_3}=\ldots=\chi_{\A_{k-2p-1}}=-1$. It is then easy
to show that the non-abelian factor of $\cg_0$ is $su(k{+}1{-}p,p)$, while the
non-abelian factor of $\ck$ is
$su(k{-}p,p)$ if $\chi_{\A_1} = -1$ and 
$su(k{+}1{-}p,p{-}1)$ if $\chi_{\A_1} = 1$.

The statements made in the last few paragraphs 
should strictly be modified very slightly for the special case 
$p=0$, since then we obviously cannot take $r=p-1$, 
and there is a unique possibility for
the centralizer $\ck = su(k) \oplus u(1)$. The fact that the result is
a compact group makes this one of the more interesting possibilities.

We now summarize what we have found for even $m \leq n$.
Only the split real form
$\cg = sl(n{+}1)$ corresponds to the identity map on the regular
subalgebra, with $\cg_0 = sl(n{-}m{+}1) \oplus m \, gl(1)$ and 
$\ck = sl(n{-}m) \oplus gl(1)$. Other allowed real forms
$\cg = su(n\!+\!1\!-\frac{m}{2}\!-\!p , \frac{m}{2}\!+\!p)$
correspond to a non-trivial diagram automorphism of the regular
subalgebra and result in
$\cg_0 = su(n{-}m{+}1{-}p,p) \oplus \frac{m}{2} gl(1) \frac{m}{2} u(1)$ 
and 
$\ck = su(n{-}m{-}r,r) \oplus u(1)$ where $r=p{-}1$ or $r=p$ (when these
values make sense).

The remaining isolated case $m=n$ admits just two possibilities
corresponding to the split form or a non-trivial diagram automorphism.

\subsubsection{An example: $(A_4,A_2)$} 

The general discussion we have just given will probably seem more
intelligible if we supplement it with a simple example. 
Consider then the regular $A_2$ subalgebra generated by the first two 
simple roots $\A_1$ and $\A_2$ of $A_4$ or, equivalently, by the 
basis vectors $e_1,\,e_2$, and $e_3$. 
The grading is given by the diagram
$$
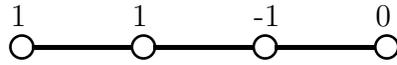
\cir{1}\li\cir{1}\li\cir{-1}\li\cir{0}
$$
and the characteristic diagram for the embedding is 
$$
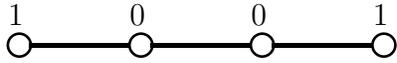
\cir{1}\li\cir{0}\li\cir{0}\li\cir{1}\,
$$
which can be found by applying a Weyl transformation to the original
basis (exchange of $e_3$ with $e_5$). From this we identify the
complex zero-grade algebra $\cg_0^{\bf c}=A_2 \oplus 2 U_1$, where
basis vectors of the root space of $A_2$ are $e_2,\,e_4$ and
$e_5$. The complex centralizer is 
$\ck^{\bf c} = A_1\oplus U_1$, where $A_1$ has root space spanned by 
$e_4$ and $e_5$. 

The possible real forms of $A_2$ are $sl(3),\,su(2,1)$ and $su(3)$. 
As always, the first can be obtained by using the trivial element of the
Weyl group. 
An element of Aut$(\DE_{A_4})$ which does not lie in the Weyl group 
must either exchange $e_4$ with $-e_5$, or else invert $e_4$ and $e_5$
individually. As shown in subsection
\ref{rfAn}, the first possibility yields $su(2,1)$, 
while the second gives rise to the compact real form $su(3)$ if we
choose $\chi_{e_2-e_4}=\chi_{e_4-e_5}=-1$. 

If we want to find which possible
real forms $\ck$ are allowed for each real form of $\cg_0$, we need
to investigate the automorphisms a little more closely. 
In this example we consider all possible choices of 
$\chi_1=\chi_{e_2-e_4}$
and $\chi_2=\chi_{e_4-e_5}$ for the automorphism which inverts $e_4$
and $e_5$. This results in the real forms:   
$$
\begin{array}{rrll}
\chi_1& \chi_2 & A^\tau_2\subset\cg_0& A^\tau_1\subset\ck\\
 -1 & -1 & su(3) & su(2) \\
  1 & -1 & su(2,1) & su(2) \\
 -1 &  1 & su(2,1) & sl(2) \\
  1 &  1 & su(2,1) & sl(2)
\end{array}
$$
The total set of possibilities can be summarized: 
$$
\begin{array}{|l|l|l|}
\hline 
\cg & \cg_0 &  \ck \\ \hline 
sl(5) & sl(3)\oplus gl(1)\oplus gl(1) & sl(2)\oplus gl(1) \\ 
su(3,2) & su(2,1)\oplus gl(1)\oplus u(1) & sl(2)\oplus u(1) \\
su(3,2) & su(2,1)\oplus gl(1)\oplus u(1) & su(2)\oplus u(1) \\
su(4,1) & su(3)\oplus gl(1) \oplus u(1) & su(2) \oplus u(1) \\ \hline 
\end{array}
$$

\subsection{Embeddings in $D_n$}
\label{Dn}

The roots of $D_n$ are 
$\Delta = \{\pm(e_k \pm e_l) \, : \,k\not=l \}$; the simple roots 
are $\alpha_i = e_i-e_{i+1}$ for $i=1,\ldots,n{-}1$ 
and $\alpha_n = e_{n-1} + e_n$. 
For $n >4$ the 
Weyl group $W(\DE)$ acts as permutations on the basis vectors $e_i$ 
together with a sign change (inversion) 
of any {\em even\/} number of them.
The group Aut($\DE$) acts as permutations 
together with a sign change (inversion) of an {\em arbitrary\/} 
number of them. This is clearly consistent with 
Aut$(\DE)/W(\DE)$ acting as a $\fl{Z}_2$ symmetry of the Dynkin diagram.
The case $n=4$ has well-known exceptional symmetry properties
(e.g.~$\mbox{Aut} (\DE) / W(\DE)$ 
is the group of permutations on three objects)
but in fact these are not relevant for our purposes. 
We shall therefore present arguments
below for the general case $n>4$; the special case
$n=4$ can be confirmed to fit into the pattern despite its
peculiarities.

\subsubsection{Real forms of $D_n$}
 
Consider an element $\tau \in \mbox{Aut}(\DE)$ for which the first 
$n-p$ basis vectors are invariant and the last $p$ are inverted. 
This gives an action on the simple roots:
\bea
\label{TD}
\begin{array}{rclr}
\tau(\A_{i}) &=& \A_{i} & i < n-p \nn
\tau(\A_{n-p}) &=& \A_{n-p} + 2 \A_{n-p+1} + \cdots + 2\A_{n-2} +
\A_{n-1} + \A_n \\ 
\tau(\A_{i}) &=& -\A_{i} & n-p < i \le n 
\end{array}
\ena
where
$\A_{n-p}+2\A_{n-p+1}+\cdots+2\A_{n-2}+\A_{n-1}+\A_n=
\A_{n-p}+\A_{n-p+1}+\psi$, and $\psi$ is the highest root of the $D_{p}$
subalgebra generated by the last $p$ basis vectors. The roots of this
subalgebra are exactly those roots which satisfy $\tau(\A) = -\A$.
For this subalgebra we can consistently choose $\chi_{\A} = -1$, and 
using eq.~(\ref{char2}) we find the character: 
\be
\label{D1} 
\sigma = (n-2p)  - 2|\DE_+(D_p)| = n-p - \dim(D_p) = n - 2p^2,
\quad p=0,\ldots,n  
\ee
where $\mbox{tr}(\tau_H) = n-2p$.
This is the character of the real form $so(n{+}p,n{-}p)$. 

In order to construct the real form $so^*(2n)$ we could consider a more
complicated element of Aut($\DE$) with 
$\frac{n}{2}$ 2-cycles if $n$ is even, and 
$\frac{n-1}{2}$ 2-cycles and one inverted basis vector if $n$ is odd. 
It is easy to check that this gives the correct real form with 
character $-n$. 
An alternative method
which will be more useful in what follows is to take 
the same root automorphism written in (\ref{TD}) above with $p=n$, 
so that all the basis vectors are 
inverted and $\tau (\alpha) = - \alpha$. From this we can define a
different conjugation map by taking $\chi_{\alpha_{n-1}}
\chi_{\alpha_n} = -1$.
Using the formula $\chi_\alpha = (-1)^{\ell + 1} \prod
\chi_{\alpha_i}^{\ell_i}$
where $\alpha = \sum \ell_i \alpha_i$ has height $\ell = \sum \ell_i$,
we find that the contributions to the character from the roots
$e_k - e_l$ 
and
$e_k + e_l$ 
always cancel for $k < l$.
The character is therefore determined solely by the action of the 
conjugation map on the Cartan subalgebra, giving the desired result 
$\sigma= - n$.

We have already seen that we can obtain the split real form $so(n,n)$
by taking $p=n$ in the construction (\ref{TD}). This is simply the
usual statement that the split real form arises from the identity in
the Weyl group.
We can also obtain the split real form from a conjugation map which 
inverts some basis vectors, however, which will be significant shortly.
To see how to do this we take the root automorphism (\ref{TD}) 
with $p=n{-}2$, so that $e_{n-1} \rightarrow - e_{n-1}$ 
and $e_{n} \rightarrow - e_{n}$ while all other basis vectors are invariant.
In contrast to what we did earlier, however, we define a different 
conjugation map by choosing $\chi_{\alpha_{n-1}} = \chi_{\alpha_n}
=1$. An easy calculation then shows that the resulting character in
$\sigma = n$, as required. 

\subsubsection{Invariant Regular Subalgebras} 

The simple regular subalgebras of $D_n$ are $A_m$ with $m<n$ and $D_m$ with 
$m\le n$. Integral embeddings arise for $A_m$ with $m$ even 
or $m=n-1$, and for all $D_m$.

\paragraph{$\bullet$ $(D_n,A_m)$.}
Consider first even $m < n{-}1$.
We can assume without loss of generality that $A_m$ is generated by the first 
$m$ simple roots of $D_n$ and 
the grading is specified by 
\be 
\label{AmDncan}
\underbrace{\cir{1}\hlih\cir{1}}_m\hspace{-.5mm}
\li\cir{$-\frac{m}{2}$}\li\cir{0}\hlih
\cir{0}\li\up{0}{0}\li\cir{0}
\ee
or in terms of the characteristic Dynkin diagram: 
\be
\label{AmDn}
\underbrace{\cir{0}\li\cir{1}\li\cir{0}\li\cir{1}\hlih\cir{1}}_m
\hspace{-.5mm}\li\cir{0}\hlih
\cir{0}\li\up{0}{0}\li\cir{0}
\ee
From this we see that the complex zero-grade subalgebra is 
$\cg_0^{\bf c} = D_{n-m} \oplus \frac{m}{2} A_1 \oplus \frac{m}{2}
U_1$ and the complex centralizer is
$\ck^{\bf c}=D_{n-m-1}\oplus U_1$.
There is exactly one basis vector which belongs to the root-spaces
of both $A_m$ and of $\cg_0^{\bf c}$, namely $e_{{m\over
2}+1}$. 

Any automorphism $\tau$ which restricts to a diagram automorphism on $A_m$ is
completely determined on $\frac{m}{2} A_1 \oplus \frac{m}{2}
U_1 \subset \cg_0^{\bf c}$ and gives the real form 
$\frac{m}{2} sl(2) \oplus \frac{m}{2} gl(1)$. This is similar to results 
which hold for the embeddings $(B_n,A_m)$ and $(C_n,A_m)$, the first
of which is discussed in appendix A; we therefore omit further details. 

Turning to $D_{n-m} \subset \cg_0^{\bf c}$, 
we can freely choose the action of the
automorphism $\tau$ on the basis vectors which generate this
subalgebra, except for the action on the 
basis vector it has in common with $A_m$. 
This basis vector must be invariant if $\tau$ leaves
$A_m$ invariant, but it must be inverted if $\tau$ acts as a
non-trivial diagram automorphism on $A_{m}$. 
From the results above, we find that from any
automorphism which acts as the identity on $A_m$ 
we can construct conjugation maps leading to the real forms 
$so(n{-}m{+}p, n{-}m{-}p)$ with $p=0,\ldots,n{-}m{-}1$. 
From an automorphism which restricts to a non-trivial diagram
automorphism on $A_{m}$ we can in fact get all these and also the
additional remaining real form
$so^*(2n{-}2m)$.
In order to show this we need to use the conjugation maps 
given earlier which involve at least one inverted basis 
vector.

The possible real forms \ck~are $so(2n{-}2m{-}2)\oplus gl(1)$ and
$so(n{-}m{-}1{+}p,n{-}m{-}1{-}p)\oplus gl(1)$ with 
$p=0,\ldots,n{-}m{-}2$ if $\tau$
leaves $A_m$ invariant, and $so^*(2n{-}2m{-}2)\oplus u(1)$ and
$so(n{-}m{-}1{+}p,n{-}m{-}1{-}p)\oplus u(1),\,p=0,\ldots,n{-}m{-}1$ if $\tau$
acts non-trivially on $A_m$. 

Finally, if $m = n{-}1$ then the grading is integral whether $n$ is odd or
even. The outstanding case $n$ even has 
$\cg_0^{\bf c}={n\over 2} A_1 \oplus {n\over 2} U_1$ and
$\ck^{\bf c}=U_1$. We simply quote the final results for the real
forms in the table on \pageref{newtable1}. 

\paragraph{$\bullet$ $(D_n,D_m)$.} 
We can assume that $D_m$ is generated by
the last $m$ simple roots of $D_n$. The grading is: 
$$       
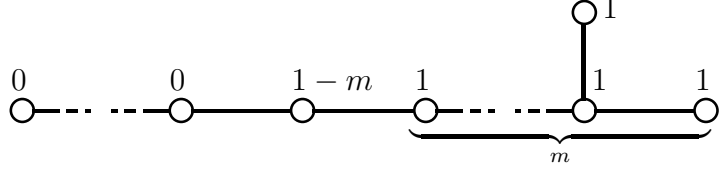
\cir{0}\hlih\cir{0}\li\cir{$1-m$}\li\hspace{-.5mm}
\underbrace{\cir{1}\hlih\up{1}{1}\li\cir{1}}_m
$$       
and the corresponding characteristic Dynkin diagram is: 
$$       
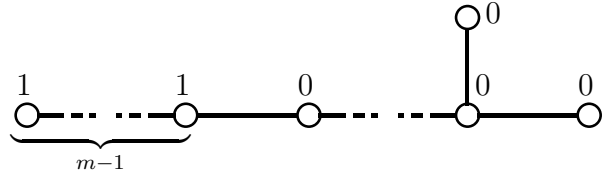
\underbrace{\cir{1}\hlih\cir{1}}_{m-1}\hspace{-.5mm}
\li\cir{0}\hlih\up{0}{0}\li\cir{0}
$$       
From this we find $\cg_0^{\bf c}=D_{n+1-m}\oplus (m{-}1) U_1$,
while the complex centralizer is $\ck^{\bf c}=D_{n-m}$. 
As in previous cases, it is important that
$e_n$ is a basis vector of the root spaces of both $D_m$ and
$\cg_0^{\bf c}$.  

An automorphism $\tau$ on $D_n$ 
which restricts to a diagram automorphism on $D_m$
will have the first $(m{-}1)$ basis vectors of $D_m$ as fixed points, while
$e_n$ is a fixed point if $\tau$ restricts to the identity on $D_m$,
but is inverted if $\tau$ is a non-trivial diagram
automorphism on $D_m$. 
We can freely choose 
the action of the automorphism on the simple part 
$D_{n+1-m}\subset\cg_0^{\bf c}$, provided that
it fulfils this criterion.

Referring back to eq.~(\ref{D1}) we see that with at least one 
invariant basis vector in $D_{n+1-m}$ we can obtain the real forms 
$so((n{-}m{+}1)+p,(n{-}m{+}1)-p)$ with $p=0,\ldots,n{-}m$. 
Alternatively, with one inverted basis vector we can get the compact
real form $so(2n{-}2m{+}2)$ and the real forms
$so((n{-}m{+}1){+}p,(n{-}m{+}1){-}p)$ with $p=1,\ldots,n{-}m$. 
For the centralizer $\ck$ we find that the possible real forms are
respectively $so(2n-2m)$ and 
$so((n{-}m){+}r,(n{-}m){-}r)$ with $r=p{-}1,\, p, \,p{+}1$. 

\subsection{Embeddings in $B_n$ and $C_n$}
\label{BnCn}
         
The cases of $B_n$ and $C_n$ are very similar to each other;
in fact the groups $W (\DE) = \mbox{Aut} (\DE)$ are isomorphic.
The roots of $B_n$ are 
$\Delta = \{ \pm e_j \, ; \, \pm(e_k \pm e_l) \, : \,k\not=l \}$ 
and the simple roots are 
$\A_i = e_i-e_{i+1}$ for $i=1,\ldots,n{-}1$ and $\A_n = e_n$. 
The roots of $C_n$ are 
$\DE = \{ \pm 2e_j \, ; \, \pm(e_k\pm e_l) : \, k\not = l\}$ 
and the simple roots are $\alpha_i = e_i-e_{i+1}$ for 
$i=1,\ldots,n{-}1$ and $\alpha_n = 2 e_n$. 
In either case the Weyl group $W(\DE)$ is the group of 
permutations of the basis vectors $e_i$ with a sign change of an arbitrary 
number of them. 
         
\subsubsection{Real Forms of $B_n$ and $C_n$} 
         
In the case of $B_n$ we consider again a conjugation map with
the first $(n-p)$ basis vectors invariant and the last $p$ inverted. 
The result is 
\beano
\begin{array}{rclr}
\tau(\A_{i}) &=& -\A_{i} & i < n-p \\
\tau(\A_{n-p}) &=& \A_{n-p} + 2 \A_{n-p+1} + \cdots + 2\A_n \\ 
\tau(\A_{i}) &=& -\A_{i} & n-p < i \le n 
\end{array}
\enano
The set of characters that we obtain via this
construction is: 
$$       
\sigma = n - p - \dim(B_p) = n - 2p(p+1), 
$$       
corresponding to the complete set of real forms of $B_n$. 

As with the case of $D_n$, it will also be important that we can
obtain the split real form from a conjugation map which inverts
one basis vector. 
The roots of $B_n$ are: 
\beano
e_k - e_l & = & \A_{k} + \cdots + \A_{l-1} \\ 
e_k + e_l & = & \A_{k} + \cdots + \A_{l-1} + 2 \A_l + \cdots + 2 \A_{n} \\
e_k & = & \A_k + \cdots + \A_n
\enano
We consider the conjugation map defined by $\A \rightarrow -\A$, 
and we choose $\chi_\A = 1$ for all $\A$. Using 
equations (\ref{chi}) and (\ref{chi2}),
a brief calculation shows that the resulting character is $\sigma =
n$, i.e the resulting real Lie algebra is indeed the split real Lie 
algebra. 
         
In the case of $C_n$ we find that any automorphism which leaves 
at least one basis vector invariant leads to the split real form with
character $n$ (details are given in appendix \ref{A1}). 
We therefore consider an automorphism consisting of $p$ 2-cycles 
acting on the first $2p$ basis vectors 
and the rest of the basis vectors inverted. 
This automorphism acts on the simple roots as: 
\beano
\begin{array}{rclr}
\tau(\A_{2i-1}) &=& -\A_{2i-1} & i \le p \\
\tau(\A_{2i}) & = & \A_{2i-1} + \A_{2i} + \A_{2i+1} & i < p \\
\tau(\A_{2p}) &=& \A_{2p-1} + \A_{2p} + 2 \A_{n-p+1} - \cdots -
2\A_{n-1} + \A_n \\
\tau(\A_{i}) &=& - \A_{i} & n-p < i \le n 
\end{array}
\enano
Choosing $\chi_\A = -1$ for the last $n-2p$ simple roots
generating a $C_{n-2p}$
subalgebra, we get the
characters:  
$$       
\sigma = -(n-2p) - 2p - 2|\DE_+(C_{n-2p})| = -2p - \dim(C_{n-2p}) = 
-n - 2(n-2p)^2
$$       
with $p = 0, \ldots , [\frac{n}{2}]$. 
These are the characters of the remaining real forms 
$sp(p,n{-}p)$ of $C_n$. 
         
\subsubsection{Invariant Regular Subalgebras} 
The simple regular subalgebras of $B_n$ are: $A_m$, $m<n$;
$B_m,\,m\le n$; and $D_m,\,m\le n$. 
For an integral embedding we require $A_m$ with $m$ even; but no
additional restrictions on $B_m$ or $D_m$.

The simple regular subalgebras of
$C_n$ are: $A_m$, $m<n$; and $C_m,\,m\le n$. 
For an integral embedding we require $A_m$ with $m$ even or 
$m = n-1$; and $C_m$ with $m=n$, giving the principal embedding.

\paragraph{$\bullet$ $(B_n,A_m)$.} 
We consider the regular $A_m$ subalgebra generated
by the first $m$ (even) simple roots of $B_n$, with grading:
$$       
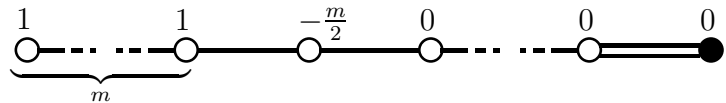
\underbrace{\cir{1}\hlih\cir{1}}_m\hspace{-.5mm}\li\cir{$-\frac{m}{2}$}
\li\cir{0}\hlih\cir{0}\dli\bul{0}
$$       
corresponding to the characteristic Dynkin diagram: 
$$       
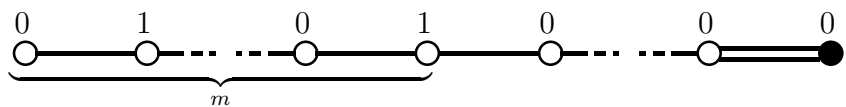
\underbrace{\cir{0}\li\cir{1}\hlih\cir{0}\li\cir{1}}_m\hspace{-.5mm}
\li\cir{0}\hlih\cir{0}\dli\bul{0}
$$       
The complex zero-grade subalgebra is 
$\cg_0^{\bf c} = B_{n-m} \oplus \frac{m}{2} A_1 \oplus \frac{m}{2} U_1$,
while the complex centralizer is 
$\ck^{\bf c} = B_{n-m-1}\oplus U_1$. 
         
Any automorphism which acts as a  diagram automorphism on $A_m$ 
will have a restriction to 
$\frac{m}{2} A_1 \oplus \frac{m}{2} U_1 \subset \cg^{\bf c}_0$ which
is completely fixed; the corresponding conjugation map defines 
the real form 
$\frac{m}{2} sl(2) \oplus \frac{m}{2} gl(1)$.
Details of this are given in appendix A.

The basis vector 
$e_{\frac{m+2}{2}}$ is common to the root-space of both $A_m$
and of $B_{n-m}\subset\cg_0^{\bf c}$. 
An automorphism which restricts to the identity
on $A_m$ must leave this basis vector invariant, while an
automorphism which restricts to a non-trivial diagram automorphism on
$A_m$ will invert $e_{\frac{m+2}{2}}$ so
we are free to choose the action of an automorphism 
on $B_{n-m}\subset\cg_0^{\bf c}$ subject only to this constraint.
Using one or the other of these possibilities we can define
conjugation maps leading to any real form of $B_{n-m}$. 
Similarly we can obtain all consistent real forms of the simple part of the
centralizer $B_{n-m-1} \subset \ck^{\bf c}$.
The abelian factor $U_1 \subset \ck^{\bf c}$ has real forms 
$gl(1)$ or $u(1)$, depending
on whether the automorphism on $A_m$ is trivial or not. 
         
\paragraph{$\bullet$ $(B_n,B_m)$.}
We consider the regular $B_m$ subalgebra generated by the last 
$m$ simple roots of $B_n$, with grading:
$$       
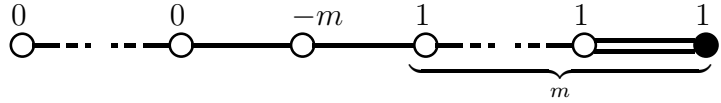
\cir{0}\hlih\cir{0}\li\cir{$-m$}\li\hspace{-.5mm}
\underbrace{\cir{1}\hlih\cir{1}\dli\bul{1}}_m
$$       
corresponding to the characteristic Dynkin diagram: 
$$       
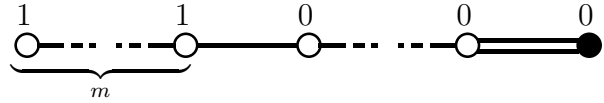
\underbrace{\cir{1}\hlih\cir{1}}_m\hspace{-.5mm}
\li\cir{0}\hlih\cir{0}\dli\bul{0}
$$       
We find $\cg_0^{\bf c}=B_{n-m}\oplus m \, U_1$ and 
$\ck^{\bf c} = D_{n-m}$. 
 
An automorphism which is trivial when restricted to $B_m$ will also be trivial
on $m U_1\subset\cg^{\bf c}_0$. 
We can, however, choose the action of the automorphism on 
$B_{n-m}\subset\cg_0^{\bf c}$ and on 
$\ck^{\bf c}$ freely, since in this case the root spaces of $B_m$ and
$\cg^{\bf c}_0$ do not intersect. 
We can therefore obtain any real form of these subalgebras.

\paragraph{$ \bullet$ $(B_n,D_m)$.} 
We consider the regular $D_m$ subalgebra generated by the first 
$(m-1)$ simple roots of $B_n$ and by the root $e_{m-1}+e_m$. 
This gives the grading: 
$$       
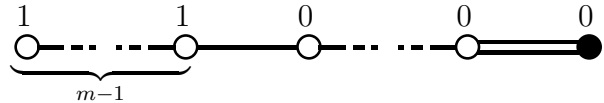
\underbrace{\cir{1}\hlih\cir{1}}_{m-1}\hspace{-.5mm}\li\cir{0}\hlih\cir{0}\dli
\bul{0}  
$$       
which is already a characteristic Dynkin diagram. 
We find $\cg_0^{\bf c} = B_{n+1-m} \oplus (m{-}1) U_1$ and 
$\ck^{\bf c}= B_{n-m}$.

The basis vector $e_m$ is common to the root-spaces of both $D_m$ and
$\cg^{\bf c}_0$.
An automorphism $\tau$ which restricts 
to a diagram automorphism on $D_m$
will leave the first $m{-}1$ basis vectors of $D_m$ invariant, while
fixing $e_m$ if $\tau$ is the
identity on $D_m$, and inverting it if it is a non-trivial diagram
automorphism. 
We can therefore choose 
the action of the automorphism on $B_{n+1-m}\subset\cg_0^{\bf c}$ 
subject only to this one constraint.
From these two classes of automorphisms we can define conjugation maps
which give any desired real form of $B_{n+1-m}\subset\cg_0^{\bf c}$ and
any consistent real form of $\ck^{\bf c}$. 

\paragraph{$\bullet$ $(C_n,A_m)$.} 
We consider first even $m<n-1$. 
The Dynkin diagram is identical to that for the reduction 
$(B_n,A_m)$ with white and black nodes exchanged.
The zero-grade subalgebra is 
$\cg^{\bf c}_0=C_{n-m} \oplus \frac{m}{2} A_1\oplus \frac{m}{2} U_1$
and the centralizer is $\ck^{\bf c} = C_{n-m-1} \oplus U_1$. 

The restriction of an automorphism to 
$\frac{m}{2} A_1\oplus \frac{m}{2} U_1\subset\cg_0^{\bf c}$ 
is completely determined when we impose that the restriction to $A_m$
must be a diagram automorphism, and the real form will be 
$\frac{m}{2} su(2) \oplus \frac{m}{2} gl(1)$ (the proof is
similar to one given in appendix A for embeddings in
$B_n$).
The basis vector $e_{{m+2\over 2}}$ is common to the root spaces of 
both $A_m$ and $\cg_0^{\bf c}$. 
The action of an automorphism on $C_{n-m}\subset\cg_0^{\bf c}$ is
constrained only by the fact that it must leave
$e_{m+2\over 2}$ invariant if it acts trivially on $A_m$, while it
must invert $e_{m+2\over 2}$ if it acts
non-trivially on $A_m$.  

Any automorphism on $C_{n-m}$ which 
leaves at least one basis vector invariant defines a conjugation map 
which results in the split real form of $C_{n-m}$.
This requires some justification and we again consign the details to
the appendix so as not to interrupt our summary. 
It follows that an automorphism which restricts to the
identity on $A_m$ can result only in the split real form of
$C_{n-m}\subset\cg_0^{\bf c}$. 
An automorphism which acts non-trivially on $A_m$ can result in
any real form of $C_{n-m}\subset\cg_0^{\bf c}$. 
We can furthermore find conjugation maps which give any consistent 
real form of $C_{n-m-1}\subset\ck^{\bf c}$. The abelian factor 
$U_1\subset\ck^{\bf c}$ always has real form $u(1)$
for an automorphism which acts as a non-trivial diagram automorphism
on $A_m$.

Finally we mention the special case $m = n-1$, 
which also gives an integer grading when $m$ is odd. 
$\cg_0^{\bf c} = \frac{n}{2} A_1 \oplus \frac{n}{2} U_1$ and the
possible real forms are $\frac{n}{2} sl(2) \oplus \frac{n}{2} gl(1)$ 
and $\frac{n}{2} su(2) \oplus \frac{n}{2} gl(1)$. 
The centralizer is $\ck^{\bf c} = U_1$ with real forms $gl(1)$ or $u(1)$. 
         
\subsection{Summary of classification} 
\label{sec-4}

\begin{theo}
The real forms $(\cg , \ch)$ compatible with an integral embedding
$(\cg^{\bf c}, \ch^{\bf c})$ where $\cg^{\bf c}$ is classical of rank
$n$ and $\ch^{\bf c}$ is simple of rank $m$ are as shown in the Table.
For each such embedding we give the complex types $\cg_0^{\bf c}$
and $\ck^{\bf c}$ of the zero-grade subalgebra and the centralizer,
followed by their corresponding real forms 
$\cg_0$ and $\ck$. We set $k=n-m$ throughout to simplify notation.
\end{theo}

\noindent
Proof: Sketched in the foregoing sections 5.1 to 5.4 with some 
additional technical details in appendix A.
\hfill $\Box$\\[3mm]

Note that in order to write the results in the most compact way we
have made free use of the standard identifications
$U_1 \cong D_1$; $A_1 \cong B_1 \cong C_1$;  
$B_2 \cong C_2$; $D_2 \cong A_1\oplus A_1$; 
$A_3 \cong D_3$ and besides the equivalence of the
real forms $sl(2) \cong su(1,1)$ and $so^*(8) \cong so(6,2)$.

\label{newtable1} 
\hspace{-25mm} 
\vbox{
$
\begin{array}{|c|c|c|}
\hline\hline 
\cg\mbox{ compatible with}&\cg_0\mbox{ real form of}&\ck\mbox{ real form of}\\
(A_n , A_m) \mbox{ $m$ even} &A_{k} \oplus m U_1 &A_{k-1} \oplus U_1\\
\hline \hline 
\vphantom{\matrix{0\cr0\cr}} 
sl(n\!+\!1) & sl(k\!+\!1)\oplus m \, gl(1) & sl(k)\oplus gl(1)\\ \hline
\vphantom{\matrix{0\cr0\cr}}  
su\left(n\!+\!1\!-\!\frac{m}{2},\frac{m}{2}\right) & 
su(k\!+\!1) \oplus \frac{m}{2} gl(1) \oplus \frac{m}{2} u(1)&
su(k) \oplus u(1) \\ \hline
\begin{array}{c} 
su\left(n\!+\!1\!-\!\frac{m}{2}\!-\!p,\frac{m}{2}\!+\!p\right) \\
p = 1, \ldots , [\hf(k{+}1)]   \end{array} 
&su(k{+}1{-}p,p) \oplus \frac{m}{2} gl(1) \oplus \frac{m}{2} u(1) 
&\begin{array}{l} su(k{-}p,p) \oplus u(1)\\
su(k{+}1{-}p,p{-}1) \oplus u(1) \end{array} \\ \hline \hline
\cg\mbox{ compatible with}&\cg_0\mbox{ real form of}&\ck\mbox{ trivial}\\
(A_n , A_n) \mbox{ $n$ odd} &n \,U_1 &\\ 
\hline\hline
\vphantom{\matrix{0\cr0\cr}} 
sl(n\!+\!1) & n \, gl(1) & \mbox{---} \\ \hline
\vphantom{\matrix{0\cr0\cr}} 
su\left({n+1\over 2},{n+1\over 2}\right) 
& \frac{n+1}{2} gl(1) \oplus \frac{n-1}{2} u(1) &
\mbox{---} \\ \hline \hline
\cg\mbox{ compatible with}&\cg_0\mbox{ real form of}&\ck\mbox{ real form of}\\
(B_n, A_m) \mbox{ $m$ even}
&B_k \oplus \frac{m}{2} A_1 \oplus \frac{m}{2} U_1 
&B_{k-1} \oplus U_1\\ 
\hline \hline 
\vphantom{\matrix{0\cr0\cr}}  
so(n\!+\!1,n)& 
so(k{+}1,k) \oplus \frac{m}{2} sl(2) \oplus \frac{m}{2} gl(1)
&\begin{array}{l}
so(k,k\!-\!1) \oplus gl(1) \\ so(k,k\!-\!1) \oplus u(1)\\
so(k\!+\!1,k\!-\!2) \oplus u(1) \end{array}  \\ \hline
\vphantom{\matrix{0\cr0\cr}}  
so(2n\!+\!1\!-\!m,m)& so(2k{+}1)\oplus \frac{m}{2} sl(2) \oplus 
\frac{m}{2} gl(1) 
&so(2k{-}1) \oplus u(1)\\ \hline
\begin{array}{c} 
so(n\!+\!1\!+\!p,n\!-\!p)\\
p=1,\ldots,k{-}2 
\end{array} 
&so(k{+}1{+}p,k{-}p)\oplus \frac{m}{2}sl(2) \oplus \frac{m}{2}gl(1)
&\begin{array}{l} 
so(k{-}1{+}p,k-p) \oplus u(1)\\ 
so(k{+}p,k{-}1{-}p) \oplus u(1)\\  
so(k{+}1{+}p,k{-}2{-}p) \oplus u(1) \end{array}  \\ \hline \hline
\cg\mbox{ compatible with}&\cg_0\mbox{ real form of}&\ck\mbox{ real form of}\\
(B_n, B_m) &B_{k} \oplus m \, U_1 &D_{k} \\
\hline \hline 
\vphantom{\matrix{0\cr0\cr}}  
so(n\!+\!1,n) & so(k\!+\!1,k) \oplus m \, gl(1)& so(k,k) \\ \hline
\vphantom{\matrix{0\cr0\cr}}  
so(2n\!+\!1\!-\!m,m) & so(2k\!+\!1) \oplus m \, gl(1)& so(2k) \\ \hline
\begin{array}{c} 
so(n\!+\!1\!+\!p,n\!-\!p) \\
p=1,\ldots,k{-}1 \end{array} 
& so(k\!+\!1\!+\!p,k\!-\!p) \oplus m \, gl(1) 
&\begin{array}{l} so(k\!+\!p,k\!-\!p) \\ 
so(k\!+\!1\!+\!p,k\!-\!1\!-\!p) \end{array} 
\\ \hline \hline
\cg\mbox{ compatible with}&\cg_0\mbox{ real form of}&\ck\mbox{ real form of}\\
(B_n,D_m) &B_{k+1} \oplus (m{-}1) \, U_1 &B_{k} \\
\hline \hline 
\vphantom{\matrix{0\cr0\cr}}  
so(n\!+\!1,n) & so(k{+}2,k{+}1) \oplus (m{-}1) \, gl(1) 
& so(k\!+\!1,k) 
\\ \hline
\vphantom{\matrix{0\cr0\cr}} 
so (2n\!+\!2\!-\!m,m\!-\!1) & so(2k\!+\!3) \oplus (m{-}1) \, gl(1) 
& so(2k\!+\!1) 
\\ \hline
\vphantom{\matrix{0\cr0\cr}} 
so (2n{+}1{-}m,m) & so(2k{+}2,1) \oplus (m{-}1) \, gl(1) 
&\begin{array}{l} so(2k,1)\\ so(2k\!+\!1) \end{array} \\ \hline
\begin{array}{c} 
so(n\!+\!1\!+\!p,n\!-\!p) \\ p=1,\ldots,k{-}1 \end{array} 
&so(k{+}2{+}p,k{+}1{-}p) \oplus (m{-}1) \, gl(1) 
&\begin{array}{l} so(k{+}p,k{+}1{-}p) \\ 
so(k\!+\!1\!+\!p,k\!-\!p) \\
so(k+\!2\!+\!p,k\!-\!1\!-\!p) \end{array} \\ \hline 
\end{array} $
}  

\hspace{-20mm} 
\vbox{
$
\begin{array}{|c|c|c|}
\hline\hline
\cg\mbox{ compatible with}&\cg_0\mbox{ real form of}&\ck\mbox{ real form of}\\
(C_n, A_m) \mbox{ $m$ even} 
&C_{k} \oplus \frac{m}{2} A_1 \oplus \frac{m}{2} U_1
&C_{k-1} \oplus U_1\\ 
\hline \hline
sr(n) 
&sr(k) \oplus  \frac{m}{2} sl(2) \oplus \frac{m}{2} gl(1) 
&\begin{array}{c} sr(k\!-\!1) \oplus gl(1)\\ 
sr(k\!-\!1) \oplus u(1)\end{array}  \\ \hline
\vphantom{\matrix{0\cr0\cr}} 
sp(\frac{m}{2},n-\frac{m}{2}) 
&sp(k)\oplus \frac{m}{2} su(2)\oplus  \frac{m}{2} gl(1) 
&sp(k\!-\!1) \oplus u(1)\\ \hline
\begin{array}{c}
sp\left ([\frac{n}{2}]\!-\!p,[\frac{n+1}{2}]\!+\!p\right ) \\  
p=1,\ldots,[\frac{k}{2}]-1 \end{array}
& 
sp\left ([\frac{k}{2}]\!-\!p,[\frac{k+1}{2}]\!+\!p \right) 
\oplus\frac{m}{2} su(2) \oplus \frac{m}{2} gl(1)
& \begin{array}{l} 
sp\left([\frac{k}{2}]\!-\!p\!-\!1,[\frac{k+1}{2}]\!+\!p \right ) \oplus u(1) \\
sp\left([\frac{k}{2}]\!-\!p,[\frac{k+1}{2}]\!+\!p\!-\!1 \right ) \oplus u(1)
\end{array}
\\ \hline \hline 
\cg\mbox{ compatible with}&\cg_0\mbox{ real form of}&\ck\mbox{ real form of}\\
(C_n, A_{n-1}) \mbox{ $n$ even}
&\frac{n}{2}A_1\oplus\frac{n}{2}U_1&U_1 \\ 
\hline \hline 
sr(n) & \frac{n}{2}sl(2)\oplus \frac{n}{2}gl(1)
& \begin{array}{c} gl(1) \\ u(1) \end{array} \\ \hline
sp\left(\frac{n}{2},\frac{n}{2}\right) 
&\frac{n}{2} su(2) \oplus \frac{n}{2}gl(1)
& u(1) \\ \hline \hline 
\cg\mbox{ compatible with}&\cg_0\mbox{ real form of}&\ck\mbox{ trivial}\\
(C_{n} , C_n) &n \, U_1 & \\
\hline \hline
\vphantom{\matrix{0\cr0\cr}} 
sr(n) & n\, gl(1) &  \mbox{---}\\ \hline\hline 
\cg\mbox{ compatible with}&\cg_0\mbox{ real form of}&\ck\mbox{ real form of}\\
(D_n, A_m) \mbox{ $m$ even}& D_{k}\oplus\frac{m}{2}A_1\oplus\frac{m}{2} U_1 
& D_{k-1} \oplus U_1 \\ 
\hline\hline
so(n,n) & so(k,k) \oplus \frac{m}{2}sl(2) \oplus \frac{m}{2} gl(1) 
&\begin{array}{l} so(k{-}1,k{-}1) \oplus gl(1) \\ 
so(k{-}1,k{-}1) \oplus u(1) \end{array}  \\ \hline
\vphantom{\matrix{0\cr0\cr}} 
so^*(2n) & so^*(2k) \oplus \frac{m}{2}su(2) \oplus \frac{m}{2} gl(1) 
& so^*(2k{-}2) \oplus u(1) \\ \hline
\vphantom{\matrix{0\cr0\cr}} 
so(2n\!-\!m,m) & so(2k) \oplus \frac{m}{2}sl(2) \oplus \frac{m}{2} gl(1) 
& so(2k{-}2) \oplus u(1)\\ \hline
\vphantom{\matrix{0\cr0\cr}} 
so(2n\!-\!m\!-\!1,m\!+\!1) 
& so(2k\!-\!1,1) \oplus \frac{m}{2}sl(2) \oplus \frac{m}{2} gl(1) 
& so(2k\!-\!3,1) \oplus u(1) \\ \hline
\begin{array}{c}
so(n\!+\!p,n\!-\!p) \\
p=1,\ldots,k{-}2 \end{array}
& so(k\!+\!p,k\!-\!p) \oplus \frac{m}{2}sl(2) \oplus \frac{m}{2} gl(1)
& \begin{array}{l} so(k\!-\!2\!+\!p,k\!-\!p) \oplus u(1) \\
so(k\!+\!p,k\!-\!2\!-\!p) \oplus u(1) \end{array} \\ \hline\hline
\cg\mbox{ compatible with}&\cg_0\mbox{ real form of}&\ck\mbox{ real form of}\\
(D_n,A_{n-1}) \mbox{ $n$ even}
&{n\over 2} A_1 \oplus {n\over 2} U_1 &U_1 \\
\hline\hline 
so(n,n) & \frac{n}{2} sl(2) \oplus \frac{n}{2}gl(1) 
& \begin{array}{l} gl(1) \\ u(1) \end{array}  
\\ \hline
\vphantom{\matrix{0\cr0\cr}} 
so^*(2n) 
& \frac{n}{2}su(2) \oplus \frac{n}{2} gl(1) 
& u(1)  
\\ \hline\hline 
\cg\mbox{ compatible with}&\cg_0\mbox{ real form of}&\ck\mbox{ real form of}\\
(D_n,D_m) & D_{k+1} \oplus (m{-}1) U_1 & D_{k} \\
\hline\hline
\vphantom{\matrix{0\cr0\cr}} 
so(n,n) 
& so(k\!+\!1,k\!+\!1) \oplus (m{-}1) \, gl(1) 
& so(k,k) \\ \hline
\vphantom{\matrix{0\cr0\cr}} 
so(2n\!-\!m\!+\!1,m\!-\!1) 
& so(2k\!+\!2) \oplus (m{-}1) \, gl(1) 
& so(2k) \\ \hline
so(2n\!-\!m,m) 
& so(2k\!+\!1,1) \oplus (m{-}1) \, gl(1) 
&\begin{array}{l} so(2k\!-\!1,1) \\ so(2k) \end{array} \\ \hline
\begin{array}{c} 
so(n{+}p,n{-}p) \\
p = 1,\ldots,k-1 \end{array} 
& so(k\!+\!1\!+\!p,k\!+\!1\!-\!p) \oplus (m{-}1) gl(1)
& \begin{array}{l} 
so(k{-}1{+}p,k{+}1{-}p) \\ so(k+p,k-p) \\ so(k\!+\!1\!+\!p,k\!-\!1\!-\!p)
\end{array} \\ \hline 

\end{array} $} 

\section{Concluding Remarks} 
\label{sec-5}

We hope the results we have described in this paper will be of
interest and of use from three related but distinct points of view.
First, for future work on integrable lagrangian field theories.
Second, for the continued investigation of $\cw$-algebras, their
representation theory, and their eventual classification. Third, 
for the purely mathematical problem of classifying all embeddings of
$sl(2,\fl{R})$ into real Lie algebras. One may anticipate, given the
wide-ranging importance of Lie algebras in many branches of
mathematics and physics, that such a classification may well have
unexpected applications in quite new areas.

There are also two more specific topics which immediately suggest
themselves.
One very interesting question is the corresponding classification 
in the case of Lie superalgebras. The significance 
of these ideas for supersymmetric models 
was originally pointed out in our earlier paper \cite{EM}, where the 
choice of the appropriate real form was shown to be essential for the correct 
realization of $N=4$ superconformal symmetry in a particular model. 
It remains to set this isolated example in a more general context,
which should also include the superconformal algebras discussed in
\cite{BiGu} as special cases.
Finally, it would also be very interesting to extend the study carried out
in this paper to the case of affine Toda theories. \\[4mm]

\noindent{\bf Acknowledgements}
 
\noindent
The research of JME is supported by a PPARC Advanced Fellowship, while  
JOM is supported by the TMR European network contract number FMRX-CT96-0012, 
and is grateful to the Danish Research Council for additional support. 
One of the authors (JOM) would like to thank 
L. Feh{\'e}r, E. Ragoucy, P. Sorba, and I. Tsutsui for useful discussions,
and Laboratoire de Physique Th{\'e}orique ENSLAPP, where much of this
work was done, for financial support
and kind hospitality. 

\appendix 
\section{Technical Results for the Classification} 
\label{A1}

In this appendix we give some 
technical details of results which were quoted in our classification 
of real embeddings in Theorem 5. 

\subsection{$D_n$} 

We will show that an automorphism on $D_n$ for $n>4$
which leaves at least one basis vector invariant 
cannot give the real form $so^*(2n)$ (note that $so^*(8)$ is
isomorphic to $so(6,2)$). 
In our proof of this we will use the character to distinguish 
real forms. 
This is not sufficient by itself for the particular case when $n=k^2$, since
the real forms $so^*(2k^2)$ and $so(n{+}k,n{-}k)$ happen to have the same
character. To complete the proof we should therefore also consider the
maximal compact subalgebra as a means of distinguishing these
real forms. We shall omit these details however.

Consider an element $\tau \in \mbox{Aut}(\DE)$ of order two,
consisting of $p$ 2-cycles, $q$ invariant basis vectors, and $r$
inverted basis vectors, so $n=2p+q+r$. 
Taking into account 
the consistency equation $\chi_\A = \chi_{\tau(\A)}$ given in 
(\ref{cons}), we find the character 
$$
\sigma = (q-r) + 2( p + \!\!\!\!\sum_{\A\in\DE_+(D_r)} \chi_\A) 
= (n-r) + (-r + 2\!\!\!\! \sum_{\A\in\DE_+(D_r)} \chi_\A ), 
$$
where $D_r$ is the subalgebra corresponding to the $r$ inverted basis
vectors. 
The term $q-r$ is the trace of $\tau_H$ and $2p$ is the contribution from the
$p$ positive roots that are inverted by the $p$ two-cycles. 
Note that, remarkably, the character depends only on $r$. 
The term $-r + 2 \sum_{\A\in\DE_+(D_r)} \chi_\A$ is the character $\sigma_r$ 
of the real form
of the $D_r$ subalgebra, which means that:
$$
-r + 2 \sum_{\A\in\DE_+^r} \chi_\A = 
\sigma_r = \left \{ \begin{array}{l} 
(r-2j^2),\,\,j= 1,\ldots,[\hf r] \\
-r \end{array} \right.
$$
To obtain the real
form $so^*(2n)$ the character should be  
$$
\sigma = (n{-}r) + \sigma_r = - n
$$
This has solutions if $n=k^2$ or if $r=n$. In the latter case, the
automorphism leaves no basis vector invariant. In the former case
we cannot distinguish, using the character alone, between 
$su^*(n)$ and $so(n{+}k,n{-}k)$. 
But consideration of the maximal compact subalgebra 
shows that it is the latter real form which is defined by this 
conjugation map.

\subsection{$B_n$} 

For the reduction $(B_n,A_m)$ discussed in section \ref{BnCn} 
we found the zero-grade subalgebra to
be of complex type $\cg_0^{\bf c}=
\frac{m}{2} U_1 \oplus\frac{m}{2} A_1 \oplus B_{n-m}$. 
We claimed that any automorphism that 
leaves $A_m$ invariant up to a diagram automorphism
will result in the real form $\frac{m}{2} sl(2) \subset \cg_0$ 
of $\frac{m}{2} A_1\subset\cg_0^{\bf c}$.  
We now justify this statement.

The roots of $\frac{m}{2} A_1$ are 
$e_1+e_{m+1},\,e_2+e_m,\dots, e_{\frac{m}{2}} + e_{\frac{m}{2}+1}$. 
Consider e.g. $e_1+e_{m+1}$. 
If $\tau$ acts as a non-trivial diagram automorphism on $A_m$ then 
$\tau(e_1+e_{m+1}) = -(e_1+e_{m+1})$ and
$\chi_{\A_1}=\chi_{\A_2}=\ldots =\chi_{\A_m}$. In order to find the
character of the real form 
of the corresponding $A_1$ we must find $\chi_{e_1+e_{m+1}}$. Considering
the roots  
$\A=\A_1+\A_2+\ldots +\A_m=e_1 - e_{m+1}$ and $\B=e_{m+1}$ we see that
$e_1+e_{m+1} = \A+2\B$. Now $\tau(\A) =\A$ and
$\tau(\B)=-(\A+\B)$, so 
$$
\chi_{\A+2\B} = \frac{N_{\A\,\B}}{N_{\A\,-(\A+\B)}} 
\frac{N_{\A+\B\,\B}}{N_{-\B\,-(\A+\B)}} \chi_\A (\chi_\B)^2 
$$
We see immediately that $N_{\A+\B\,\B}=N_{-\B\,-(\A+\B)}$. 
Also $\chi_\A =-1$ because 
$$
\A = (\A_1+\cdots+\A_{\frac{m}{2}})+(\A_{\frac{m}{2}+1}+\cdots+\A_m)
\equiv \B_1 + \B_2
$$
with $\tau(\B_1)=\B_2$. By symmetry we have 
$\chi_{\B_1} = \chi_{\B_2}$, so  
$\chi_\A = \frac{N_{\B_1\,\B_2}}{N_{\B_2\,\B_1}}
\chi_{\B_1}\chi_{\B_2}= -1$. 
Finally, $N_{\A\,\B}= - N_{\A\,-(\A+\B)}$ because 
$$
N_{\A\,\B}N_{\A\,-(\A+\B)} = N_{\A\,\B} N_{\A+\B\,-\A} = \A \cdot \B =
-1
$$
where we have used the Jacobi identity. We conclude that 
$\chi_{e_1+e_{m+1}} = 1$ and we can now calculate the character of the 
corresponding $A_1$ to be 1, i.e. the character of $sl(2)$.

Similar (but slightly more complicated) calculations can be 
done in the cases of $(C_n,A_m)$ and $(D_n,A_m)$ referred to in the text. 

\subsection{$C_n$}

We show that any conjugation map on $C_n$ with at least one 
invariant basis vector leads 
to a real form with character $\sigma = n$, i.e.~to the split real Lie
algebra. We will limit ourselves to show this in the case where the
automorphism has only invariant and inverted basis vectors. The
calculation when the automorphism also contains 2-cycles is not
significantly different. 

Consider a conjugation map on $C_n$ which leaves the 
first $p$ basis vectors invariant and which inverts the remaining 
$n{-}p$ basis vectors.
The action of $\tau$ on the simple roots is: 
\beano
\begin{array}{rclr}
\tau(\A_{i}) &=& \A_{i} & i < p \\
\tau(\A_{j}) &=& -\A_{j} & j > p \\
\tau(\A_{p}) = \tau(e_p - e_{p+1}) &=& e_p + e_{p+1} 
\end{array}
\enano
Setting $\gamma = e_{p + 1} - e_n$ and writing 
$$
e_p+e_{p+1}=(e_p-e_{p+1})+ (e_{p+1}-e_n)+(e_{p+1}-e_n)+2e_n=
\A_p + \G + \G + \A_n
$$
we get the consistency condition 
$$
\chi_{\A_p} = \chi_{e_p+e_{p+1}} = 
\frac{N_{\A_p\,\G}}{N_{\A_p+2\G+\A_n\,-\G}}
\frac{N_{\G\,\A_n}}{N_{-\G\,-\A_n}} \chi_{\A_p} \chi_{\A_n}=
\chi_{\A_p}\chi_{\A_n} 
$$
since $N_{\A_p\,\G}/N_{\A_p+2\G+\A_n\,-\G}=
N_{\G\,\A_n}/N_{-\G\,-\A_n}=-1$. This implies 
$\chi_{\A_n}=1$. 

Corresponding to the inverted basis vectors, there
is a subalgebra of type $C_{n-p}$ 
with roots satisfying $\tau(\A) = -\A$, and these can be written in the form: 
\beano
e_k - e_l & = & \A_{k} + \cdots + \A_{l-1} \\ 
e_k + e_l & = & \A_{k} + \cdots + \A_{l-1} + 2 \A_l + \cdots + 2 \A_{n-1} 
+ \A_n \\
2e_k & = & 2\A_k + \cdots + 2\A_{n-1} + \A_n
\enano
This, together with the equation (\ref{chi}), shows that when $\chi_{\A_n}=1$ 
the contribution to the character from the roots $e_k-e_l$ cancels 
the contribution from the roots $e_k+e_l$, and that 
$\chi_{2e_k} = \chi_{\A_n}$. The total character is then: 
$$
\sigma = (p-(n-p)) + 2 (n-p)  = n 
$$
where the first term arises from tr$(\tau_H)$, while the 
second term is the contribution from the roots 
$2e_{p+1},2e_{p+2},\ldots,2e_n$. \\[4mm]
\vfill \eject

\section{Calculating Toda Potentials}

In this appendix we indicate how to calculate the non-abelian 
Toda potentials  
$$
V = \mbox{tr} (M_+ g M_-g^{-1})
$$
for the cases of non-principal embeddings in $B_2$ and $G_2$
discussed in section 4. The general idea is simply to identify the 
representations of $G_0$ in which $M_\pm$ transform under 
conjugation in $G$, since this is what occurs in $V$ above.
The actual calculations are particularly easy to carry out when the
only non-abelian factors in $G_0$ are of the simple type $A_1$.

Consider the spin-1/2 representation of $sl(2)$ 
acting on a pair of states labelled $\pm$ to indicate their
eigenvalues $\pm 1/2$ under the diagonal generator.
With a standard choice of coordinates $\phi_0$ and $\phi_\pm$ we have 
$$
g \equiv
\pmatrix{
U_{++} & U_{+-}\cr
U_{-+} & U_{--}\cr
}
=
\exp \pmatrix{ \phi_0 & \phi_+ \cr \phi_- & - \phi_0 \cr}
=
{1 \over \rho} \pmatrix{ \rho \cosh \rho + 
\phi_0 \sinh \rho & \phi_+ \sinh \rho\cr
\phi_- \sinh \rho & \rho \cosh \rho - \phi_0 \sinh \rho \cr}
$$
where $\rho = \phi_0 + \phi_+ \phi_-$.

In the case of $B_2$ the generators $M_\pm$ transform in the spin-1 
representation, which we may regard as a symmetrized product of two
spin-1/2 representations. The relevant  
matrix acting on states labelled by $+1$, 0, $-1$ is
then 
$$
\pmatrix{ U^2_{++} & U_{++}U_{+-} & U_{+-}^2 \cr
U_{++} U_{-+} & \frac{1}{2} (U_{++}U_{--}+U_{+-}U_{-+}) & U_{+-} U_{--} \cr 
U^2_{-+} & U_{-+}U_{--} & U_{--}^2 \cr
}
$$
Taking the $(B_2,A_1)$ description of the embedding, we have $M_\pm$ 
appearing as states of spin 0. 
The contribution to the potential is therefore given by the
`centre' element of this matrix:
$$
(U_{++}U_{--}+U_{+-}U_{-+}) = {1 \over \rho^2} ( \rho^2 \cosh^2 \rho
- \phi_0^2 \sinh^2 \rho + \phi_+ \phi_- \sinh^2 \rho )
= { 1 \over \rho^2} ( \phi_0^2 + \phi_+ \phi_- \cosh 2 \rho )
$$
as given in the text. The potential for the $(B_2, A_1 \oplus A_1)$
embedding can be calculated similarly, but involves the components
with spins $\pm 1$ instead.

In the case of $G_2$ the generators $M_\pm$ transform in the spin-3/2
representation, with states labelled by $+3/2$, $+1/2$, $-1/2$,
$-3/2$.
Once again we can construct this as the 
totally symmetrized product of three spin-1/2 representations. 
The matrix in question is 
$$
\pmatrix{ U^3_{++} & U^2_{++}U_{+-} & U_{++}U_{+-}^2 & U_{+-}^3 \cr
U^2_{++} U_{-+} 
& \frac{1}{3} (U^2_{++}U_{--}+ 2U_{+-}U_{++}U_{-+}) 
& \frac{1}{3} (U^2_{+-}U_{-+}+ 2U_{+-}U_{++}U_{--}) 
& U^2_{+-} U_{--} \cr 
U_{++} U^2_{-+} 
& \frac{1}{3} (U^2_{-+}U_{+-}+ 2U_{-+}U_{++}U_{--})  
& \frac{1}{3} (U_{++}U^2_{--}+ 2U_{+-}U_{--}U_{-+})  
& U_{+-} U^2_{--}\cr
U^3_{-+} & U_{-+}^2 U_{--} & U_{-+} U^2_{--}  & U^3_{--}\cr
}
$$
For the $(G_2, A_2)$ embedding the elements $M_\pm$ involve the states
of extreme spin $\pm 3/2$.
It is easy to see that the contribution to the potential is therefore 
given by the sum of the `corner' entries in the matrix above:
$$
U^3_{++} + U^3_{--} + U^3_{+-} + U^3_{-+} 
= {1\over\rho^3} ( \, 2 \rho^3 \cosh^3 \rho 
+ 6 \rho \phi_0^2 \cosh \rho \sinh^2 \rho +
\phi^3_+ \sinh^3 \rho + \phi^3_- \sinh^3 \rho \, ) 
$$
where $\rho = \phi_0^2 + \phi_+ \phi_-$, which is again the result 
given in the text.


\begin{thebibliography}{99}
\bibitem{LS} A.N. Leznov and M.V. Saveliev, \CMP{89} (1983) 59;
\CMP{74} (1980) 11.
\bibitem{ORetc} L. Feh{\'e}r, L. O'Raifeartaigh, P. Ruelle, I. Tsutsui, 
{\it Ann.~Phys.}~{\bf 213} (1992) 1;
J. Balog, L. Feh{\'e}r, L. O'Raifeartaigh,
P. Forg{\'a}cs and A. Wipf, {\it Ann.~Phys.}~{\bf 203} (1990) 76.
\bibitem{BTD} F.A. Bais, T. Tjin and P. van Driel, \NPB{357} (1991)
632.
\bibitem{PhysRep} L. Feh{\'e}r, L. O'Raifeartaigh, P. Ruelle, I. Tsutsui, 
A. Wipf, Phys.~Rep.~{\bf 222} (1992) 1.
\bibitem{BS} P.~Bouwknegt and K.~Schoutens, Phys.~Rep.~{\bf 223}
(1993) 183.
\bibitem{EM}J.M. Evans and J.O. Madsen, \PLB{384} (1996) 131,
hep-th/9605126. 
\bibitem{BoHaTj}J. de Boer, F. Harmsze, T. Tjin, 
Phys. Rep {\bf 272} (1996) 139. 
\bibitem{Dy} E. Dynkin, Amer. Math. Soc. Transl. Ser. 2, 6 (1957) 111.
\bibitem{BiGu} B. Bina and M. G\"unaydin, \NPB{502} (1997) 713;
hep-th/9703188. 
\bibitem{FRS} L. Frappat, E. Ragoucy and P. Sorba, 
\CMP{157} (1993) 499.  
\bibitem{U1} J.O. Madsen and E. Ragoucy, \CMP{185} (1997) 509, 
hep-th/9503042. 
P. Sorba, \PL{B279} (1992) 319. 
\bibitem{BoTj} J. de Boer and T. Tjin, \CMP{160} (1994) 317; 
\CMP{158} (1993) 485. 
\bibitem{FeOrIt} P. Bowcock and G.M.T. Watts, \NPB{379} (1992) 63;
L. Feh{\'e}r, L. O'Raifeartaigh, and I. Tsutsui, \PLB{316} 
(1993) 275.
\bibitem{Hel} S. Helgason, {\it Differential Geometry, Lie Groups, and 
Symmetric Spaces} (Academic Press, New York, 1978). 
\bibitem{SW}A.A.~Sagle and R.E.~Walde, {\it Introduction to Lie Groups
and Lie Algebras} (Academic Press, New York, 1973).
\bibitem{Ba} J. Balog, L. O'Raifeartaigh, P. Forg\'acs and A. Wipf,
\NPB{325} (1989) 225; 
S. Hwang, \NPB{354} (1991) 100, \PLB{276} (1992) 451;
I. Bars, \PR{D53} (1996) 3308; 
J.M. Evans, M.R. Gaberdiel and M.J. Perry, {\it The No-ghost Theorem
for $AdS_3$ and the Stringy Exclusion Principle}, 
hep-th/9806024, to appear in {\it Nuclear Physics B}.
\bibitem{BH}E. Witten, \PR{D44} (1991) 314; 
J-L. Gervais and M.V. Saveliev, \PLB{286} (1992) 271;
S. Chaudhuri, J.D. Lykken, \NPB{396} (1993) 270;
A. Bilal, \NPB{422} (1994) 258. 
\bibitem{JE} J.M. Evans, \NPB{390} (1993) 225.
\bibitem{Hon} A. Honecker, \NPB{400} (1993) 574.
\bibitem{PB} P. Bowcock, {\it Representation Theory of a W-algebra
from Generalized DS Reduction}, preprint DTP 94-5, hep-th/9403157.
\end{thebibliography}
\end{document}